\documentstyle[aps,pre,epsf,twocolumn]{revtex}
\begin{document}
\def\be {\begin{equation}}
\def\ee {\end{equation}}
\def\bee {\begin{eqnarray}}
\def\eee {\end{eqnarray}}
\def\N {{\cal N}}
\def\z {\zeta}
\def\zk {\zeta_k}
\def\OP {\tensor P}
\def\B.#1{{\bbox{#1}}}
\renewcommand{\thesection}{\arabic{section}}
\title{{\rm Submitted to J.Stat.Phys.,    \hfill  July 26,2002}
\\~~
\\Pattern Selection:  Determined by Symmetry and Modifiable by Long-Range Effects}
\author {Mitchell J. Feigenbaum}
\address{The Rockefeller University, 1230 York Ave. New York, NY 10021}
\maketitle
\begin{abstract}
We consider Saffman-Taylor channel flow without surface tension on a high-pressure
driven interface, but modify the usual infinite-fluid in infinite-channel 
configuration.  Here we include the treatment of efflux by considering a finite 
connected body of fluid in an arbitrarily long channel, with its second free 
interface the efflux of this configuration.  We show that there is a uniquely 
determined translating solution for the driven interface, which is exactly the 
1/2 width S-T solution, following from correct symmetry for a finite channel flow.
We establish that there exist no perturbations about this solution corresponding to
a finger propagating with any other width: Selection is unique and isolated.  The 
stability of this solution is anomalous, in that all freely impressible 
perturbations are stabilities, while unstable modes request power proportional to 
their strength from the external agencies that drive the flow, and so, in 
principle, are experimentally controllable.  This is very different from the 
behavior of the usual infinite fluid: The limit of infinite length is singular, 
and not that of the literature. We argue that surface tension on the {\em efflux} 
interface modifies channel-width $\lambda$ according to $1-2\lambda=\sigma/v$
(i.e. $(2\pi)^2 B$ of the 
literature) with $v$ the velocity of the high-pressure tip, but $\sigma$ the surface 
tension of the efflux.  That is, $\lambda$ is decreased below $1/2$ by the long-range 
effect 
of smoothing an arbitrarily distant efflux.  The perturbation theory created here 
to deal with transport between two free boundaries is novel and dependent upon a 
remarkable symmetry.
\end{abstract}
\section{Introduction}

Objects arising from flows, aggregation and so forth often have distinctive shapes.  
(Think, for example, of snowflakes.)  Upon theoretical study, (the identification of 
all forces and processes at work, and the determination of all boundary conditions), 
the solution will be that of this particular shape.  In other cases, however, the 
usual form is just one of many possible solutions that can be found.  The question of 
what then distinguishes and determines only the usual, observed form is termed a 
problem of ``pattern selection".  Clearly something has been left out of the 
theoretical treatment, which means that after a careful treatment, something apt to be 
subtle remains to be dealt with.  ``Left out" itself is subtle, since the theoretical 
considerations, and their solutions, entail approximations, and the ``truth" can have 
been left out by the approximations.  We consider such a problem here.

Perhaps the earliest studied selection problem arose in the 1958 work of Saffman-
Taylor (S-T) \cite{58ST}.  Here the question is the shape of a bubble penetrating into a body 
of viscous fluid.  Loosely, a soda-straw is filled with a thick fluid which is then 
forcibly blown out.  How has the breath of air propagated down the straw, to drive out 
what amount of the fluid?  Performed carefully, the straw of length $l$ is flattened 
down all along its length into a long rectangle of width $w$ and length $l$, where 
$2w$ is just about the circumference of the cylindrical straw, and a channel of 
thickness $b \ll w$ all that remains of the third dimension of the straw's volume.  
The material of the ``straw" is transparent glass for visualization.  That is, there 
are two $w$ by $l$ plates of glass, one a gap $b$ above the other, and the long edges 
are sealed. View the $w$ by $l$ rectangles from above with the long length horizontal 
(the $x$-axis) and the width $w$ along the $y$-axis.  A controlled source of high 
pressure air is now applied to the left end (say, at $x$=0), and the driven viscous 
fluid flows out the right end into atmosphere (the efflux).  The bubble of air 
advancing from the left into the viscous fluid, the immiscible high-pressure interface 
between the driving air and viscous fluid, hereafter referred to as $H$, is the object 
of study.  (See Fig.1, but imagine the viscous fluid to continue beyond $L$.)

\narrowtext
\begin{figure}
\epsfxsize=8.5truecm
\epsfbox{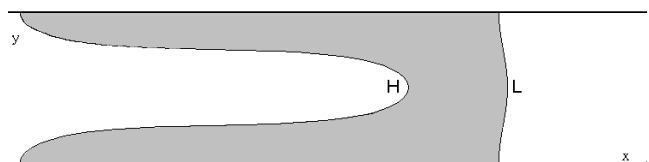}
\caption{The channel in $(x,y)$ coordinates.  The body of viscous fluid is 
shaded. It is bounded between a high-pressure interface $H$ with ``air"
and another, $L$ at low-pressure.}
\end{figure}

Invariably, S-T observed the bubble to develop into a very long ``finger", that is, a 
rounded nose furthest to the right with long horizontal sides parallel to the long 
sides of the channel trailing back towards the high pressure end.  This finger is 
almost perfectly centered within the channel, that is, with center-line symmetry.  
Moreover, S-T discovered that with higher and higher pressures, and hence increasingly 
faster bubbles, the width of the horizontal sides of the finger approached $1/2$ the 
width $w$ of the channel.  This is the selected pattern for this problem.

S-T then theoretically determined $H$ of finger shape, but of width 
$\lambda w$ with $\lambda$ anywhere from 0 to 1, rather than just $1/2$.  They had no 
argument as for why only the observed value should occur, and so exposed a 
problem of pattern selection.

Let us be more careful and say more about the experiment.  There is no reason to use 
just air to drive the viscous fluid.  Indeed, any fluid immiscible with the fluid 
initially filling the channel may be employed to drive it.  The problem is simplest if 
the driving fluid, like air, has a negligible viscosity, so that pressure is spatially 
uniform throughout it, and hence all along the interface $H$.  However, there is 
always some surface tension along $H$, so that when $H$ is curved, {\em just} within 
the viscous fluid along $H$, pressure is {\em not} uniform.  It is straightforward to 
see (say on dimensional grounds) that this surface tension $\sigma$ can modify the 
problem only in the combination $\sigma /v$ with $v$ a velocity, say the velocity of 
the tip of the finger.  It then follows that the selection of $1/2$, occurring for 
high velocities, is selection in the limit $\sigma /v \rightarrow 0^+$.  This then 
sets up the question of how this {\em singular} limit accounts for the observed 
selection.  It is singular because, as determined in the 1958 paper, with 
$\sigma \equiv 0$, there is no selection at all.

There is an extensive literature, culminating in the mid-1980s and employing
``beyond all orders" of perturbation methodology to verify this singular limit, and 
hence the critical role played by even the slightest of surface tensions acting along 
the boundary of the selected pattern.\cite{86Shr,86HL,86CDHPP}  
It is now the accepted wisdom that all these 
indeterminate selection problems achieve their theoretical resolution through 
additional ``constituentive" relationships (such as $\sigma$ for S-T) imposed directly 
upon the shape to be so selected.

Let us again be more careful, and say more about the theory.  The gap $b$ is viewed as 
so small, that with a usual limit of the viscous fluid equations, this ``film" of 
fluid can be treated purely within 2-D.  This is a theoretically compelling 
circumstance, since the problem then rapidly is noticed to be fully amenable to 
complex, conformal mathematics. \cite{87Tan,86How} 
Apart from some studies as to how finite values of
$b$ can modify the 2-D solution, the entire literature on S-T flow is 2-D conformal.

The theoretical and experimental literature nevertheless diverge in an obvious way.  
While discussing surface tension on the bubble interface, the experimental literature 
is relatively vague about how the viscous fluid emerges from the apparatus at low 
pressure.  It is not that this is totally unimportant - care has to be paid to 
terminations, and the reliable reproducibility of fast long fingers emerges only after 
some adjustment of the efflux termination.  Perhaps it is largely undocumented in 
consequence of an implicit belief that far enough away from the observed interface, 
these details should significantly have decoupled.

The theoretical literature has gone much further.  The {\em entire} literature 
considers only $l=\infty$, namely, an infinitely long channel all filled with the 
viscous fluid.  This is evidently a very different geometry from that of the 
experiments that determine selection.  Granted that subtleties can potentially lead to 
selection or its absence, one might wonder that this approximation to very large 
length has an impact upon selection.  This question has never been asked in the 
literature.  We do so here, and indeed discover that the limit $l \rightarrow \infty$ 
is also singular.

Since conformal mathematics is so powerful, we consider a finite termination of 
geometrically simplest form.  Namely, we allow the channel containing the fluid to be 
arbitrarily long, while the viscous fluid, always within it, is of a fixed finite 
volume.  Abutting the fluid at its low pressure, right end, we again have air; that is, 
a fluid with negligible viscosity, as is the high pressure driving fluid.  Hence, we 
now have two interfaces; the high-pressure, experimentally observed one $H$, and the 
low-pressure, efflux one at the right, which we now term $L$.  Both interfaces are 
dynamically free surfaces, although they are evidently highly coupled, since between 
them resides the connected body of our incompressible viscous fluid.  (See Fig.1)

In this finite configuration, the pressure drop across the viscous fluid becomes 
physically finite, whereas it is simply infinite in the usual treatment.  There is 
also a net flux of fluid.  In the infinite treatment it freely can be set at any value 
with impunity, and is conventionally set at the constant 1.  In our finite case, the 
impedance of the flow, the ratio of net pressure drop by flux is again finite and 
large, as opposed to simply infinite in the usual treatment.  More importantly, this 
impedance generally changes in time, since the closest distance between the two 
interfaces dynamically changes in time, and generally decreases as the interface $H$ 
grows more bent into a finger.  In experiment Ref.\cite{87TZL}, the time derivative of 
impedance is 
indeed significantly negative.  Although impedance variations are buried in an 
infinite value in the usual infinite treatment, it is troublesome that there is no 
trace of its derivative in that literature.

The reason we have considered a finite body of fluid with definite pressures along its 
free boundaries is that there is a mathematical clue to anticipate an influence upon 
shape.  The theoretical works on the infinite fluid all consider periodic boundary 
conditions for the cross-channel ($y$) behavior.  The channel has two long sides a 
distance $w$ apart, one edge at $y$=0 and another at $y$=$w$. Interface $H$ (the only 
one for the infinite problem) has viscous fluid everywhere to its right in the 
direction of increasing $x$ (and hence decreasing pressure, $p$, with 
$p \rightarrow -\infty$ as $x \rightarrow +\infty$).  The literature takes $p$ 
periodic in $y$ with period $w$ so the $x$-values of $H$ at $y$=0 and $y$=$w$ agree.  
The edges are impenetrable with no separation of fluid from them, so that the velocity 
of fluid along them is purely in the $x$-direction, and so, $y$=0 and $y$=$w$ are 
streamlines for the flow.  Periodicity of $w$, in synchronizing the flow at both edges 
is clearly a higher symmetry than required. It is readily seen that this implies 
center-line symmetry, with $v$ also in the $x$-direction along $y$=$w/2$.  This, 
however, is just the symmetry of the observed finger,  and so the symmetry in which to 
seek a theoretical result.  This then is the choice of the literature.

Implicit in this choice of symmetry is, of course, the underlying fact that the flow 
is along the $x$-direction at each wall.  This has a striking consequence not noticed 
in the previous literature.  The conformal machinery utilizes an invertible analytic 
mapping $\zeta =h(z)$ from the fluid's spatial plane $z \equiv x+i y$ to a new plane
$\zeta \equiv -p+i s$ with $p$ pressure, and $s$ the ``stream function", where the 
curves of constant $s$ are always orthogonal to those of constant pressure.  The thin 
limit of the viscous fluid equations is Darcy's law:  The velocity is the negative 
gradient of pressure.  Thus, fluid flows with a velocity at a point, at each moment of 
time, in the direction of a curve of constant $s$ through that point at that time.  
(These curves all evolve in time as then does $h$.)  In particular, with velocity in 
the $x$-direction along the walls, this is to say real in the complex description, 
since Darcy's law reads $v_x-i v_y=h'(z)$.  This implies that at each instant of time 
the part of the strip between $y$=0 and $y$=$w$ all to the right of $H$ is mapped by 
$h$ into a strip between $s$=0 and $s$=$w$ all to the right of $p$=0 by the obvious 
choice of the origin of pressure (only differences matter).  But then, the real 
$z$-axis is mapped by $h$ to the real $\zeta$-axis.  This, in consequence of analytic 
continuation, implies that $h$ is Schwarz reflection symmetric, 
$\bar{h}(\bar{z})=h(z)$.
(This, together with an identical statement for the upper 
edge implies periodicity over $2w$, the true symmetry of the flow.  Increasing the 
symmetry to $w$ periodicity is equivalent to an extra reflection symmetry through 
$y$=$w/2$.)

The equation of motion for the interface $H$ is that it flow into itself under its 
velocity at each instant as determined by Darcy's law.  Expressed in the 
$\zeta$-plane, this becomes a relationship along $H$ (which is just Re$\zeta$=0) of 
both the $\zeta$ and time derivatives of $f=h^{-1}$ and their complex conjugates.  
With $f$ reflection symmetric, these conjugates are then just the functions themselves 
evaluated at the conjugate of $\zeta$.  But, along $H$, the conjugate of $\zeta$ is 
simply $-\zeta$.  Thus, along $H$ the equation of motion is an analytic partial 
differential equation in $\zeta$ and time.  But then it analytically continues as that 
same equation as a field equation throughout fluid and its continuation.  \cite{93Tan,1} 
That is, 
there is a new symmetry of the equations of motion, {\em parity} in $\zeta$.  The 
symmetry of $s \rightarrow -s$ is the periodicity of the literature.

Unnoticed in the literature, but implicitly present, is the symmetry of 
$p \rightarrow -p$.  That is, the equations of motion express a relationship between 
positive and negative values of $p$.  This is extraordinary, since it expresses some 
connection between the shape of $H$, determined by singularities in $f$ at (positive) 
high pressures to behaviors of the fluid at (negative) low pressures, {\em i.e.} to 
its efflux. This paper explores that connection, and discovers it be replete with 
consequences.

In particular, for a propagating finger, $\lambda$=$1/2$ is uniquely determined for 
the finite geometry with pressure fixed on each interface just within the viscous 
fluid, and so with zero surface tension on both $H$ and $L$.  This immediately implies 
that the limit $l \rightarrow \infty$ is singular, and S-T pattern selection is 
unambiguously resolved in finite geometry.

The class of S-T solutions for arbitrary $\lambda$ in the infinite problem is 
particularly simple.  When we consider how they transport fluid particles arbitrarily 
far downstream, it is easily discovered that only the $\lambda$=$1/2$ solution has 
cross-channel curves of particles all at a fixed value of pressure all flowing into 
other curves of fixed pressure at later times.  With no surface tension, the fluid may 
be opened up to atmosphere along any such curve, which then is $L$.  The actual form 
of the $\lambda$=$1/2$ solution came from mathematical simplicity in the infinite 
problem.  In the finite version it is the {\em unique} solution to the question of a 
translating finger.  The other S-T solutions for $\lambda < 1/2$ far downstream (far 
to the right) correspond to solutions for the finite problem with $H$ still without 
surface tension, but with $\sigma /v=1-2\lambda$ on $L$, the curve that can then be 
opened to atmosphere.  This confirms the significant connection between the shape of 
$H$ and some behavior (smoothing) of $L$.  Moreover only $\lambda < 1/2$ is physically 
allowed, but requiring enormous values of $\sigma$ to substantially decrease $\lambda$ 
from $1/2$.  For viscosities and high velocities of the experimental literature, this 
amounts to no more than a several percent reduction in $\lambda$.  Reference \cite{87TZL}
found 
just such an order of magnitude reduction from $1/2$, which there was interpreted as 
the absence of any genuine selection, and speculated that it might be in consequence 
of films of fluid left behind on the glass plates behind the advancing finger.  
Moreover, it violates the singular limit theory of $\sigma \rightarrow 0^+$, which 
precisely determines just $1/2$.  Finiteness can possibly solve this puzzle in the 
literature.

Next, with $\sigma$=0 on $H$ for the infinite literature, all the S-T solutions are 
linearly unstable.  The $\sigma \rightarrow 0^+$ theory stabilizes its $\lambda$=$1/2$
solution.  The finite problem without surface tension, significantly, has its 
$\lambda$=$1/2$ solution stable in a somewhat anomalous sense.  When we consider
fluctuations about a given solution, we presume that they can be imposed with an 
arbitrarily small expenditure of power.  In the finite problem the same instabilities,
that in the $\sigma$=0 infinite geometry grew exponentially, require power exactly 
proportional to their size in order to be sustained.  If there is no source to provide 
this power, then the solution is fully stable in finite geometry.  This means that if the
pump driving the flow is prevented, say by some feedback control, from engaging in rapid
variations in the power requested from it, then the solution is stabilized.  This is so
different from the infinite geometry, that we again realize $l \rightarrow \infty$ is
singular.  The $l \rightarrow \infty$ limit of the $\sigma \rightarrow 0^+$ theory has
never been worked out, let alone questioned.  It clearly should be.

The idea that the treatment of an arbitrarily distant efflux can modify the shape of an
object has never been experimentally checked.  It is worth bearing in mind that
aggregation arises in diffusive processes for which again there is an incompressible
fluid with identical dynamics - namely the probability flux of random walkers.  There
is enough sufficiently delicate in the considerations to follow in this paper, that one
can only go so far as to speculate that quite to the contrary of the accepted wisdom,
patterns might more generally depend on details remote from the selected shape itself.

This paper is laid out as follows:

In Section 2. we erect the conformal field equations suitable to two free
interfaces, and notice that they are displacement invariant under a time $\varphi$,
the integrated flux up to time $t$.  In order to determine exact solutions to 
perturb about, we ask the natural questions:  Are there purely translating
solutions, and are there solutions which
automatically satisfy the equations of {\em both} interfaces, simply in
consequence of this translation invariance?  

In Section 3. we establish that the 
only such solutions turn out to be the 1/2 width Saffman-Taylor fingers, here not
guessed at, but uniquely determined.  We determine that Saffman-Taylor fingers of
a narrower width correspond to gathering and smoothing the efflux in a precise
way, and no matter how distant that efflux is from the driven interface.

Section 4. develops the machinery to explicitly solve the finite problem in
perturbation theory.  A felicitous symmetry of the equations of both interfaces allows
us to fully integrate both equations into a purely algebraic problem.  Although it is 
known that the infinite case is fully integrable, that is not known in the finite
geometry we discuss.  This symmetry allows one to speculate that this problem too is
completely integrable.
A result of this effort is that all
unstable modes, and only these, determine (unstably growing) external impedances,
and this one function, just of time, uniquely encodes the entire unstable spatial
flow, thus allowing a pump driving the flow to ``hear" the precise shape of the
interface simply by detecting the power requested from it, and so, rendering it
capable of completely stabilizing the flow.  No such thing is true in the infinite
case, and so we realize that just how the flow is terminated matters, no
matter how long the body of fluid.  

In Section 5. we realize that while formal perturbative pole-dynamical solutions
\cite{45Pol,45Gal,94DMW,84SB}
that apparently produce different width fingers exist, these solutions
are {\em purely} formal, and can't correspond to any actual approximate solutions
in the finite problem nearby to the 1/2 finger.  This fully establishes isolated
pattern selection.

The Discussion Section 6. more iconically presents what this paper has accomplished
after the reader has been armed by the machinery presented.

The paper ends in an Appendix that establishes that the
pure S-T 1/2 finger is unique, and so totally isolated within pole-dynamics for the
finitely terminated problem.  It has been relegated
to an appendix in order not to interrupt the main line of exposition.
This Section contains some modest new results for 
pole-dynamics, and is a self-contained extension to paper \cite{1}.  That paper can also 
assist the reader in gaining familiarity with the workings of our reflection symmetric 
field equations.  Indeed, some of the material of this Section was reflected into that
paper. 
 
As a final point of information, in each numbered Section, {\em e.g.} 2., equation numbers 
start again from 0, and are referred to within just that same Section without a Section 
number prefix.  References
to equations in Sections external to a given Section are then prefixed by the number of 
the Section they appear in.


\section{The General Setup and Equations of Motion}

Taking $z=x+iy$, with $x$ increasing down-stream along a channel of cross-width
$y\in[0,\pi]$, the fluid obeys Darcy's law for the velocity $v$
and pressure $p$
\begin{equation}
v=-\partial p \ . \label{1}
\end{equation}
The flow is incompressible so that
$\partial\cdot v=0=\Delta p$, and so $p$ is
harmonic, and the real part of an analytic function $h$, naturally
constructed through its derivative:
\begin{eqnarray}
h' &\equiv& \frac{\bar{v}}{V(t)}=\frac{v_x-iv_y}{V(t)}=
\frac{-\partial_x p
+i\partial_y p}{V(t)}\nonumber \\
h &\equiv& \xi +i\eta=\frac{-p+i s}{V(t))}=\zeta \ ,\label {2}
\end{eqnarray}
which satisfies Cauchy-Riemann for $\Delta p=0$, and $s$ is the harmonic 
conjugate of $-p$. $V(t)$ here is a function of time alone. Consider the
change in $h$ along an arbitrary curve $\gamma$ 
\begin{equation}
 \delta h =\int_\gamma h' dz=-\frac{1}{V}\delta p +
  \frac{i}{V}\int_\gamma v_n d \ell \ . \label{3}
\end{equation}
Since by Eq.(\ref{2}) $\delta\xi=-\delta p/V(t)$,
$$\delta \eta=\frac{\pi}{V(t)}\left(\frac{1}{\pi}\int_\gamma v_n d \ell
\right)$$
and so
\begin{equation}
 \delta \eta \equiv\pi~{\rm for}~V(t) =  \frac{1}{\pi}\int_\gamma v_n
d \ell
\ . \label{4}
\end{equation}
Thus with $V(t)$ the mean velocity of all fluid (the conserved flux
divided by cross-width $\pi$), the original spatial channel is
analytically mapped to an identical one in $\zeta$. (Although flux is
conserved throughout the incompressible fluid, for a finite system
it will generally vary in time.)

Assuming the fluid not to stagnate other than in front of added
boundaries, $v \neq 0$, $h' \neq 0$, and so $h$ is invertible,
{\em conformally} mapping one [0,$\pi$] strip onto another.  We
denote its inverse as $f$:
\begin{equation}
f=h^{-1}, f'=\frac{V}{\bar{v}} \; \; {\rm{at}} \; \; z= f(\zeta)\
\ . \label{5}
\end{equation}

The virtue of $f$ is that all boundaries are known in $\zeta$: The
problem we consider has impenetrable walls at $\eta = 0,\pi$ (along
which $v$ is real) and two free interfaces, one at $p \equiv 0$, i.e.,
$\xi=0$ and the other at $p=p_{atm} \equiv -p_g$, or at
\begin{equation}
\xi = \xi_g \equiv p_g/V \ . \label{6}
\end{equation}
$p_g>0$ is the gauge pressure across the body of fluid, and $\xi_g$
(the total impedance) positive and generally time-dependent, because
either or both of $p_g$ and $V$ are. ($p_g$ is time dependent if we
experimentally modulate the driving pressure.) Thus, in $\zeta$, the
fluid is just a rectangle extending from $\xi=0$ up to a variable
right hand edge.  

To track a fluid particle through use of the
one-parameter (time) family of maps, $f$, tag a particle at $t=0$ by
its
$\zeta$-coordinate $\zeta_0$:
\begin{equation}
z(\zeta_0,t)=f(\zeta(\zeta_0,t), t); \; \; \; \zeta (\zeta_0,0)
\equiv \zeta_0 \ . \label{7}
\end{equation}
Then
\begin{eqnarray}
v=z_t = f'\zeta_t+f_t & = & V/\bar{f'}, \; \; {\rm{or}} \nonumber
\\
V & = & |f'|^2 \zeta_t + f_t \bar{f'} \ . \label{8}
\end{eqnarray}

Once $f$ and $V$ are known, Eq.~(\ref{8}) is an ordinary
differential equation to
be solved for $\zeta(\zeta_0,t)$ given the initial data of Eq.(\ref{7}).

To determine $f$ requires the physical input that an interface simply
transports under its velocity. On an interface ${\rm Re}\zeta=\hat{\xi}(t)$,
with $\hat{\xi}=0$ on the left interface and $\hat{\xi}=\xi_g(t)$
on the right. That is, if ${\rm Re} \zeta_0 = \hat{\xi}(0)$ then
${\rm Re} \zeta(\zeta_0,t) = \hat{\xi}(t)$ for a fluid particle on the
interface.
That is,
\begin{equation}
{\rm Re} \zeta_t=\dot{\hat{\xi}}(t) \; \; {\rm{on}} \; \;
{\rm Re} \zeta=\hat{\xi}(t) \ , \label{9}
\end{equation}
so that (\ref{8}) is
\begin{equation}
V(t)={\rm Re} (\bar{f'}(f'\dot{\hat \xi}+f_t)) \; \; {\rm{on}} \; \;
{\rm Re} \zeta = \hat{\xi} \ , \label{10}
\end{equation}
on each interface.  The form of (10) recommends the following
definition:
\begin{equation}
f_{\hat{\xi}}(\zeta,t)\equiv f(\zeta+\hat{\xi}(t), t)\ , \label{11}
\end{equation}
immediately yielding 
\[
\hspace{15mm}
V(t)={\rm Re}(\bar{f}'_{\hat{\xi}} f_{\hat{\xi},t}) \; \; {\rm{on}}
\; \; {\rm Re} \zeta=0
\hfill (10')
\]
on each interface ${\rm Re} \zeta = \hat{\xi}(t)$.

The generally unknown-in-advance $V(t)$ complicates $(10')$. Delaying its
determination to a final step, it is overwhelmingly expedient to now
eliminate it through a redefinition of time:
\begin{equation}
\dot{\varphi}\equiv V \ , \label{12}
\end{equation}
so that $\varphi$-time is integrated flux to time $t$, that is, the volume of 
fluid moved, or the energy
expended by external sources in the case $p_g \equiv const$.  But then
$\partial_t = V \partial_\varphi$ and $(10')$ is simply 
\[
\hspace{17mm}
1 = {\rm Re}(\bar{f}'_{\hat{\xi}} f_{\hat{\xi},\varphi}) \; \; {\rm{on}} \;
\; {\rm Re} \zeta=0 \ . 
\hfill (10'') 
\]
Finally since $f$ takes ${\rm Im} \zeta=0$ to ${\rm Im}z=0$, $f$ is
Schwarz-reflection symmetric,
\begin{equation}
f(\bar{\zeta},t)=\overline{f(\zeta,t)} \ . \label{13}
\end{equation}
But ${\rm Re}\zeta = 0 \rightarrow \bar{\zeta} = - \zeta$, and $(10'')$
reads
\begin{equation}
2 =
f'_{\hat{\xi}}(-\zeta)f_{\hat{\xi},\varphi}(\zeta)+f'_{\hat\xi}(\zeta)
f_{\hat{\xi},\varphi}(-\zeta) \ . \label {14}
\end{equation}

Although (\ref{14}) is exactly $(10'')$ on ${\rm Re}\zeta=0$,
(\ref{14}) is in fact correct over
all of $f$'s analytic continuation, since (\ref{14}) is an equation just of the
analytic variable $\zeta$ and (\ref{14}) and $(10'')$ agree
all along the curve of an interface.

For our problem of two free interfaces, let us call $f_{\hat{\xi}=0}$
simply $f$, and $f_{\xi_{g}}\equiv g$:
\begin{equation}
g(\zeta,\varphi) = f(\zeta + \xi_g(\varphi),\varphi) \ . \label {15}
\end{equation} 

Then the equations of motion (\ref{14}) are simply 
\[
\hspace{15mm}
2=f'(-\zeta)f_\varphi(\zeta)+f'(\zeta)f_\varphi(-\zeta) \ ,
\hfill (14')
\]
\[
\hspace{15mm}
2=g'(-\zeta)g_\varphi(\zeta)+g'(\zeta)g_\varphi(-\zeta) \ .
\hfill (14'') 
\]
That is, $f$ and its $\xi_g$ translate $g$ are {\em both}
solutions to $(14')$.  Although $(14')$ is strange in its $\zeta$-dependence,
it is simply autonomous (translation-invariant) in $\varphi$.  That is,
\begin{equation}
f(\zeta,\varphi) \;  {\rm{a \; solution \; to}} \; (14')\Rightarrow 
f(\zeta, \varphi + \varphi_0) \;  {\rm{also \; a \; solution}}  .
\label {16}
\end{equation}

This is significant in that purely translating solutions are allowed 
and a path is also opened for {\em both}
$f$ and $g$ to obey $(14')$ in a way independent of the matching-up of
inner details, merely reliant on translation invariance.

Pause to notice that reflection symmetry for $f$ has determined a new symmetry,
parity, $\zeta\rightarrow -\zeta$, for the equations of motion $(14')$.  This 
entails not only a cross-channel symmetry, but also a relation of $p$ to $-p$.
But then, far upstream singularities which determine the shape of the driven
interface become related to far downstream properties, which is to say how the
flow is terminated.  This observation is the foundation for the work that follows.

Not only is $f$ reflection symmetric about the $\xi$-axis, but also about 
Im$\zeta=\pi$ since the upper horizontal wall there is also impenetrable.  Coupled
with $f$'s reflection symmetry, this is simply the periodicity of $f$:
$$f(\zeta+2\pi i)=f(\zeta)+2\pi i.$$

Reflection symmetry itself means that both the channel and its $\zeta$-image may
be mentally reflected through the $x$ and $\xi$-axes respectively, so that, in light
of the above,
both views are periodic in $y\in [-\pi,\pi]$, although only the upper half is
physical.  However, the experimental flow actually {\em is} very close to
symmetric about its center
line.  We now capitalize on this by taking the physical channel to be $[-\pi,\pi]$,
so that reflection symmetry now enforces actual symmetry, and later, if we so choose,
study the stability of this symmetry.

That is, our channel is now of width $2\pi$ and Eq.(\ref{4}) is for the sequel
modified to
$$ V(t) =  \frac{1}{2\pi}\int_\gamma v_n d \ell \ . $$

\section{ Pattern Selection and Physical Finiteness}
\setcounter{equation}{0}

Since the equations for the interfaces are $\varphi$-translation invariant,
within an available infinite 
channel we can seek solutions for which the driven $(p=0)$ interface retains its
shape, but merely translates 
in  $\varphi$-time. By analytic continuation, such an $f$ must be of form
\begin{equation}
f=\beta(\varphi)+u(\zeta). \label{2.1}
\end{equation}
Then, by $(2.14')$
\begin{equation}
\frac{1}{\beta '(\varphi)}=\frac{u'(\zeta)+u'(-\zeta)}{2}\equiv \lambda \label{2.2}
\end{equation}
for $\lambda$ some real constant. Neglecting any branching considerations and 
taking $u$ analytic at $\zeta=0$, (\ref{2.2}) integrates to
\begin{equation}
\frac{u(\zeta)-u(-\zeta)}{2}= \lambda \zeta \label{2.3}
\end{equation}
for the antisymmetric part of $u$. On $\xi=0$ , with $u$ reflection symmetric
$${\rm{Im}}\; u(i \eta)=y(\eta)=\lambda \eta$$
so that the driven interface occupies a fraction $\lambda$ of the channel 
width. Thus $\lambda \le 1$. With 
$\lambda < 1$ the interface must stretch off to minus infinity. By ``finite" we 
shall then request that at least 
the right interface (the ``efflux") be of bounded $x$-extent. These solutions 
with $\lambda < 1$ are called 
fingers in consequence of their long asymptotically horizontal sides at 
$\pm i \pi \lambda$. (We will 
determine truly finite configurations at the end of this section.)

By (\ref{2.2}), (\ref{2.3})
\begin{equation}
f=\frac{\varphi}{\lambda}+u(\zeta) \label{2.4}
\end{equation}
with
\begin{equation}
u(\zeta)= \lambda \zeta +E(\zeta) \; ; \; E(-\zeta)=E(\zeta). \label{2.5}
\end{equation}
Clearly $E$ must have a discontinuity of $2 \pi i(1-\lambda)$ throughout 
fluid to meet channel wall data, 
so that $E$ must contain logarithms. Otherwise neither $\lambda$ nor $E$ is 
determined.

Next
\begin{equation}
g=\frac{\varphi}{\lambda}+u(\zeta+\hat{\xi}(\varphi)) \label{2.6}
\end{equation}
must again satisfy $(2.14')$:
\[
2=(u'_+ + u'_- )/\lambda +2\hat{\xi}'u'_+ u'_- \; {\rm with} \;
 u'_\pm =u'(\pm\zeta+\hat{\xi}).
\]
Multiplying by $2 \lambda^2/u'_+ u'_-$ and rearranging,
\begin{equation}
F(\hat{\xi}+\zeta)F(\hat{\xi}-\zeta)=1+4\lambda ^2 \hat{\xi}'(\varphi) \label{2.7}
\end{equation}
with
\begin{equation}
F(\zeta) \equiv 2\lambda /u'(\zeta)-1 . \label{2.8}
\end{equation}

Substituting (\ref{2.8}) into the second of (\ref{2.2}),
\begin{equation}
F(\zeta)F(-\zeta)=1 \label{2.9}
\end{equation}
so that
\begin{equation}
F(0)=\pm 1 .    \label{2.10}
\end{equation}
Setting $\zeta=\hat\xi$ in (\ref{2.7}), at which it must be defined,
\begin{equation}
F(\hat{\xi}+\zeta)F(\hat{\xi}-\zeta)=\pm F(2\hat{\xi}) \label{2.11}
\end{equation}
and
\begin{equation}
1+4\lambda ^2 \hat{\xi}'=\pm F(2\hat{\xi}) . \label{2.12}
\end{equation}

There are two possibilities: $\hat\xi$ varies with $\varphi$ or is just a constant. 
If $\hat{\xi}'=0$ then (\ref{2.7}) and (\ref{2.9}) become
$$F(\zeta+2\hat\xi)=F(\zeta)$$
and hence $u'$ is doubly periodic with real period $2\hat\xi$ and imaginary period 
$2\pi i$. But then, by (\ref{2.5}) $E'(\zeta)$ is an 
odd such doubly periodic function. Not to be just constant (and hence zero), 
$E'$ must have singularities in 
each cell. They cannot lie in $\xi \in (0,\hat\xi)$ where $f$ must be analytic.
But then, they can't lie in $(\hat\xi,2\hat\xi)$ either, since 
then, by periodicity they lie in $(-\hat\xi,0)$ and then by oddness, in 
$(0,\hat\xi)$ again. Thus singularities must appear on 
either $\xi=0$ or $\xi=\hat\xi$. With $\lambda < 1$, there must be simple poles to 
pick up the requisite $2 \pi i(1-\lambda)$ discontinuity. With no net 
integral around a cell, the sums of the residues must vanish, and so there must 
be at least two poles per cell. 
Finally, the cut in $E$ must run the length of fluid, and so there must be 
poles on both $\xi=0$ and $\xi=\hat\xi$. But then, 
the right interface must also be unbounded, which we have rejected by 
finiteness. That is, we have
\begin{equation} 
\lambda=1, \; f=\varphi+\zeta \label{2.13}
\end{equation}
or $\hat{\xi}' \neq 0$.

If $\hat{\xi}' \neq 0$, the only continuous solutions to (\ref{2.11}) with $\hat\xi$ and 
$\zeta$ independently varying is exponential:
\begin{equation} 
F(\zeta)=\pm e^{-n \zeta} \label{2.19}
\end{equation}
with $n$ an integer to meet $2\pi i$ periodicity. But then, 
$$u'=\frac{2\lambda}{1\pm e^{-n \zeta}},$$
\begin{equation}
u=\frac{2\lambda}{n} \ln (e^{n \zeta}\pm 1)=
        2\lambda (\zeta+\frac{1}{n}\ln (1\pm e^{-n \zeta})). \label{2.20}
\end{equation}

Thus, $n \geq 1$ (fluid to the right of $\xi$) {\em and} $\lambda=1/2$, so that
\begin{equation}
f=2\varphi+\zeta+\frac{1}{n}\ln (1\pm e^{-n \zeta}) \label{2.21}
\end{equation}
and wall boundary geometry is satisfied, since the coefficient of $\zeta$ is 1. 
$n>1$ is simply $n$ parallel fingers
each of width $1/2 n$, and for 
simplicity, and the one finger case we care about from experiment,
\begin{equation}
f=2\varphi+\zeta+\ln (1+ e^{-\zeta}) \label{2.22}
\end{equation}
where we have chosen the + to have the nose of the finger at channel center. 
This is {\em precisely} the 1/2 
width Saffman-Taylor finger. However, both $\lambda=1/2$ and the precise form of $E$ 
are here totally and uniquely 
determined. That is, with the efflux physically treated (the right interface at 
atmospheric pressure) pattern 
selection is complete. (The great bulk of this paper consists in comprehending 
its stability.)

Returning to (\ref{2.12}) to determine $\hat\xi$, 
$$\hat{\xi}'=-1+e^{-2n\hat{\xi}}$$
or, 
$$e^{2n\hat{\xi}}=1+k e^{-2n\varphi}$$
or, with $n=1$,
\begin{equation}
e^{2\hat{\xi}}=1+k e^{-2\varphi} \label{2.23}
\end{equation}
where $k$ determines, at reference time $\varphi=0$, the length $L_0$ of fluid from 
the center of the driven interface at 
$\ln 2$ to the center of the right interface at $\ln (e^{\hat{\xi}(0)}+1)$, or
\begin{equation}
k=4 e^{2L_0}(1-e^{-L_0}).  \label{2.24}
\end{equation}

Before further considering same details of this solution, and rendering the 
solution truly finite, let us 
consider what we have accomplished, and why efflux treatment determines 
selection. We proceed in two 
steps. First, we consider a class of ``solutions" containing (\ref{2.22}) in order to 
see just how sharply $\lambda=1/2$ is 
selected, and how it is contingent upon the treatment of efflux. In the 
bulk of this paper that follows, 
we consider if (\ref{2.1}), exact translation, isn't simply too precise and perhaps 
not quite physical a request. 
(We shall vindicate it.)

By (\ref{2.22}) and (\ref{2.5}), $$E=\ln 2 \cosh \zeta /2,$$
and so $(2.14')$ for just the driven interface is satisfied by 
$$f=\varphi/\lambda + \lambda \zeta +2(1-\lambda)E $$
or,
\begin{equation}
f=\varphi/\lambda + \zeta +2(1-\lambda)\ln (1+e^{-\zeta}), \label{2.25}
\end{equation}
for any $\lambda \in (0,1)$. These are the Saffman-Taylor fingers for width 
$\lambda$. As we already know, there is no $\xi_g$ 
for any $\lambda \neq 1/2$ for which $g$ will also satisfy $(2.14')$. Let us see why.

To track fluid particles under (\ref{2.25}), we determine $\zeta (\zeta_0,\varphi)$ by 
(2.8):
\bee
(\xi '+i\eta ')\left| 1+(2\lambda -1)e^{-\zeta}\right| ^2 \nonumber \\
=(\frac{1}{\lambda}-1)
\left( e^{-2\xi}-1+2 i e^{-\xi}\sin \eta \right). \label{2.26}
\eee
It is immediately clear that the equation decouples in $\xi$ if and only if 
$\lambda=1/2$, in which case $$\xi '=e^{-2 \xi}-1$$
(just (\ref{2.23}), of course) and fluid particles all with the same value of $\xi$ at 
time $\varphi$, but with all 
different $\eta$ values, flow again at later times into all with the same value 
of $\xi$. This is so no matter 
how far downstream we look. And just so, for $\lambda \neq 1/2$, no matter how far 
downstream, a curve of fluid 
particles all at the same $\xi$  at time $\varphi$ will fail to be so at any 
later time. That is, by opening the 
channel to the atmosphere all along the curve $\xi=\hat{\xi}(\varphi)$ for 
$\lambda=1/2$, the flow will continue to remain at atmospheric 
pressure. However for $\lambda \neq 1/2$, this flow may {\em never}, no matter how 
far downstream, be able to be opened to atmosphere.

But with $f$ differing from free flow by $O(e^{-\xi})$, which for large $\xi$ is 
astronomically smaller than any 
physically measurable entity, perhaps our question of $\lambda=1/2$ is purely 
mathematical, and not physical. Let us 
now see that it is highly physical, with measurable consequences, indeed 
consequences that conceivably have already 
been experimentally measured, and in the literature.

By (\ref{2.26}) $$\frac{d\xi}{d\eta}=- \frac{\sinh \xi}{\sin \eta}$$
and so,
\begin{equation}
\tanh \frac{\xi}{2} \tan \frac{\eta}{2}=\tan \frac{\theta}{2} \label{2.27}
\end{equation}
for that curve $\zeta(\varphi)$ which at $\xi \rightarrow +\infty$ ends at $\eta=\theta$. 
(\ref{2.27}) has the series development
\be
\eta=\theta + \sum_{1} \frac{2}{n}e^{-n\xi}\sin n\theta . \label{2.28}
\ee
We proceed now to solve (\ref{2.26}) just far downstream, retaining terms just of 
$O(e^{-\xi})$:
\be
\xi '(1+2(2\lambda -1)e^{-\xi}\cos \eta )\sim -(\frac{1}{\lambda}-1).  \label{2.29}
\ee
By (\ref{2.28}) to $O(1)$, $$\cos \eta \sim \cos \theta$$
and integrating, 
$$\xi-2(2\lambda -1)e^{-\xi}\cos \theta \sim \xi_0 (\theta)-(\frac{1}{\lambda} -1)\varphi
\equiv \hat{\xi}(\varphi,\theta) \; \xi_0 \gg 1$$
and so,
\be
\xi \sim \hat{\xi}(\varphi,\theta)+2(2\lambda -1)e^{-\hat{\xi}}\cos \theta , \label{2.30}
\ee
and by (\ref{2.28}),
\be
\eta \sim \theta +2 e^{-\hat{\xi}}\sin \theta .    \label{2.31}
\ee
Next, to $O(e^{-\xi})$, 
$$z=f \sim \zeta +2(2\lambda-1)e^{-\zeta}+\varphi/\lambda$$
and so, substituting (\ref{2.30}), (\ref{2.31}),
\be
z \sim \xi_0 +\varphi +i\theta +2\lambda e^{-\hat{\xi}+i\theta} , \label{2.32}
\ee
and at each time $\varphi$, $\theta$ parameterizes a curve of fluid particles spanning the 
channel. For $\xi_0$ large enough, 
at $\varphi=0$ these are just a flat interface by selecting those fluid particles with 
$\xi_0(\theta) \equiv \xi_0$ independent of $\theta$.

Let us now compute the curvature of $z(i \theta)$. This is just, with the center of the 
curvature to the left, $$\kappa=\frac{1}{|z'|}{\rm{Re}} \left(\frac{z''}{z'}\right).$$
With $z'=1+2 \lambda e^{-\hat{\xi}+i \theta}$, to $O(e^{-\xi})$,
\be
\kappa=2\lambda e^{-\hat{\xi}}\cos \theta \label{2.33}
\ee
and so, (\ref{2.30}) reads
\be
\hat{\xi}(\varphi) \sim \xi +(\frac{1}{\lambda}-2)\kappa . \label{2.34}
\ee

Consider now {\em treating} the efflux, namely endowing the right efflux interface 
with a surface tension $\sigma$, and 
maintaining the pressure against this interface constant along it. Then, 
$$\hat{p}(\varphi)=p-\sigma \kappa$$
and dividing by $-V(\varphi)$, $$\hat{\xi}(\varphi)=\xi+\frac{\sigma}{V}\kappa .$$

But, this is precisely (\ref{2.34}) with
\be
\frac{1}{\lambda}-2=\frac{\sigma}{V} , \label{2.35}
\ee
and this result no matter how asymptotically far downstream we open the flow to 
atmosphere. Now, $f'$ of (\ref{2.25}) at $\zeta=0$ (the nose of the finger) is $\lambda$, 
and so, the mean $V$ is related 
to the finger's speed $v$ by $V=\lambda v$. Then (\ref{2.35}) is
\be
1-2\lambda=\frac{\sigma}{v}\equiv (2\pi)^2 B \sim 40B \label{2.36}
\ee
with the dimensionless $B$ that of the literature
$$B^{-1}=12 \frac{\mu v}{\sigma}\left( \frac{w}{b}\right)^2$$
with $\mu$ viscosity, $w/b$ the ratio of channel width to the gap between the upper 
and lower plates, and $2\pi$ 
comes from our natural choice of channel width, rather than 1.

We now see that with $\sigma>0$, we can only produce fingers of width smaller than 
$1/2$ by physical smoothing 
and ``gathering" means, and that the surface tension must be physically 
enormous to significantly decrease 
$\lambda$ from $1/2$. For the physical choices of $\sigma$ at experimental conditions when 
one observes $\sigma/v$ so small as to 
be close to zero surface tension, one has $B \sim 10^{-3}$. Had this been our configuration, 
with no surface tension on the 
driven interface, but this value for the efflux interface, we would produce a 
$\lambda \sim .48$. Because we now realize 
that how the {\em efflux} is treated does matter, and since we have no idea just 
how it was accomplished in 
the published literature (because the experimentalists presumed it didn't matter),
all we can say is that we 
have explained both qualitatively and to within order of magnitude a puzzle in 
the literature\cite{87TZL}.

Just as importantly, we have determined that the infinite channel-full of fluid 
is physically an incomplete 
specification of an experiment. In never physically treating efflux, the pole at 
$\rm{Re}\; \zeta \rightarrow +\infty$ of (\ref{2.25}) promiscuously 
allows the physical completion of an experiment in any way whatsoever. For a 
given physical method of 
collecting efflux, the fluid can be opened to air at any large, but
nevertheless, finite length down the 
channel. This limit can then be taken to infinity, but for example, in our 
configuration with a second 
interface, leaving intact the selection of (\ref{2.36}). That is, the pole at 
infinity is actually, and physically 
illegally, within fluid, and so the usual theory of the infinite channel is too 
compliant to diverse stresses.

Since finiteness matters, it is useful to determine a finite equivalent to 
(\ref{2.1}). We can utilize $\varphi$-invariance 
more deeply by requesting that the second interface is none other 
than the first, but delayed in time by $\varphi_0$:
\be
f(\zeta+\hat{\xi}(\varphi,\varphi_0),\varphi)=f(\zeta,\varphi-\varphi_0)+\beta(\varphi_0).
\label{2.37}
\ee
By this we mean first that a constant, $\beta(\varphi_0)$ is allowed since $(2.14')$ entails 
just derivatives. Clearly, the amount 
of delay is related to the length of the body of fluid, so that the impedance $\hat\xi$ 
at the right interface depends 
upon it as well. We seek solutions for any chosen length of fluid, and so for 
arbitrary $\varphi_0$. Finally, 
we seek a shape of the driven interface that, in the spirit of (\ref{2.1}) is 
independent of just what length we 
contemplate so that $f$ itself is independent of $\varphi_0$. As the length of 
the body of fluid goes to zero, 
as with $\hat\xi$ of (\ref{2.23}) with $k \rightarrow 0$, we request
\be
\hat{\xi}(\varphi,0)\equiv 0 \; \Rightarrow \; \beta(0)=0. \label{2.38}
\ee
Let us now determine the solutions to (\ref{2.37}).

Set $\varphi_0=\varphi$ in (\ref{2.37}) and define
$$\hat{\xi}(\varphi,\varphi) \equiv \zeta_*(\varphi) \; ; \; f(\zeta,0) \equiv u(\zeta).$$
then, $$f(\zeta+\zeta_*(\varphi),\varphi)=\beta(\varphi)+u(\zeta)$$
or,
\be
f(\zeta,\varphi)=\beta(\varphi)+u(\zeta-\zeta_*(\varphi)). \label{2.39}
\ee
Substituting in (\ref{2.37}), with 
$$\gamma(\varphi,\varphi_0) \equiv \hat{\xi}(\varphi,\varphi_0)-\zeta_*(\varphi)+
\zeta_*(\varphi-\varphi_0),$$
obtain
$$u(\zeta+\gamma(\varphi,\varphi_0))-u(\zeta)=\beta(\varphi-\varphi_0)-\beta(\varphi)
+\beta(\varphi_0) .$$
Differentiating on $\zeta$, should $\gamma$ vary with $\varphi,\varphi_0$ then 
$u' \equiv const$, and $u=\zeta$. Otherwise, with $\gamma$ constant, setting 
$\varphi_0=0$, by (\ref{2.38}) $\gamma \equiv 0$, in which case 
$\beta(\varphi-\varphi_0)=\beta(\varphi)-\beta(\varphi_0)$ and so 
$\beta(\varphi)=\varphi/\lambda$ for some constant $\lambda$. Thus,
\be
f(\zeta,\varphi)=\frac{\varphi}{\lambda}+u(\zeta-\zeta_*(\varphi)) \label{2.40}
\ee
and
\be
\hat{\xi}(\varphi,\varphi_0)=\zeta_*(\varphi)-\zeta_*(\varphi-\varphi_0). \label{2.41}
\ee

We now ask that (\ref{2.40}) satisfies $(2.14')$. But this is precisely the discussion 
for $g$ of (\ref{2.6}) upon replacing 
$\hat{\xi}(\varphi)$ by $-\zeta_*(\varphi)$, but without (\ref{2.9}) so that 
$F(0) \equiv a$ is unknown. Certainly $\zeta_* ' \neq 0$, since 
otherwise $\hat{\xi} \equiv 0$ by (\ref{2.41}). So, by (\ref{2.20}),
$$u(\zeta-\zeta_*)=2\lambda (\zeta-\zeta_* +\frac{1}{n}\ln (1+a e^{-n(\zeta-\zeta_*)}))$$
with $a e^{-n(\zeta-\zeta_*)}\equiv \pm e^{-n(\zeta-\hat{\zeta})}$ and $\hat{\zeta}<0$, 
we must again have $\lambda=1/2$ for channel geometry.  (\ref{2.12}) is now 
$$\zeta_* '=1-a^2 e^{2 n \zeta_*}$$
or, 
$$\hat{\zeta}'=1-e^{2n\hat{\zeta}}$$
or,
$$e^{-2n\hat{\zeta}}=1-e^{-2n\varphi}$$
together with 
\be
f=2\varphi+\frac{1}{n}\ln(e^{n(\zeta-\hat{\zeta})}\pm 1).  \label{2.50}
\ee
But this is just the simplest pole-dynamical $1/2$ finger, converging to (\ref{2.21}) as 
$\varphi \rightarrow +\infty$. Indeed, (\ref{2.21}) is the limit 
of these finite solutions as the length $\varphi_0 \rightarrow +\infty$ {\em and} we take 
$\varphi \rightarrow +\infty$, but leaving a finite length of fluid downstream.

Thus, with $n=1$ and the + sign, and in the above limit of a truly finite 
problem, we have precisely two solutions related to translation:
\[
\hspace{3mm}
\lambda=1,\; f=\varphi+\zeta,\; \hat{\xi}=L_0\equiv const,\; e^{2\hat{\xi}}\equiv k
\hfill (\ref{2.13})
\]
and
\[
\hspace{3mm}
\lambda=1/2,\; f=2\varphi+\ln(e^{\zeta}+1),\; e^{2\hat{\xi}}=1+k e^{-2\varphi}. 
\hfill (\ref{2.22})
\]
For a fully finite version, we have
\bee
\lambda=1/2,\; f=2\varphi+\ln(e^{\zeta-\hat{\zeta}}+1),\; e^{-2\hat{\zeta}}=1+e^{-2\varphi},
\nonumber
\\ \hat{\xi}=\hat{\zeta}(\varphi)-\hat{\zeta}(\varphi-\varphi_0), \;
e^{2\hat{\xi}}=\frac{1+k e^{-2\varphi}}{1+e^{-2\varphi}}. \label{2.42}
\eee
with
$$\varphi_0=L=A/2\pi,\; A={\rm area \; of \; fluid,\; }k=e^{2\varphi_0}.$$

\section{ Reflection Symmetric Perturbation Theory For Two Interfaces}
\setcounter{equation}{0}

Having asked, in the light of  experiment, for purely translating solutions, we 
must now answer the 
question if the fluid in this theory of no surface tension actually assumes the 
selected solution, or instead, 
even when initially set, deeply unstably runs away from them. There is a related 
question to this. For $L_0 \gg 1$ in (3.\ref{2.24}) and $\varphi \ll L_0$, by 
(3.\ref{2.23})
$$\hat{\xi}\sim L_0-\varphi +\ln 2$$
and $\hat{\xi}$ together with the length of fluid downstream from the driven interface 
are very large. By the time 
there has been a net flow of fluid $\sim L_0,\; \hat{\xi}\sim 1$ and the right interface 
has begun to bend into a 1/2 finger parallel to 
the left interface, with the distance between the two interfaces rapidly turning 
into a thin film for larger 
values of $\varphi$. It is intuitive that a long body of fluid downstream can help 
stabilize the driven interface, but 
certainly not just a film of fluid. That is, one can expect qualitatively 
different behaviors with $\varphi \ll L_0$ and $\varphi \simeq L_0$ . 
Let us call this former regime the ``finger regime". This raises a new question. 
Is it possible that during the 
finger regime we have solutions exponentially close (i.e. $\sim e^{-L_0}$) to purely 
translating, but then rapidly, for $\varphi \sim L_0$ 
becoming highly dynamical. Then, from the physical viewpoint of an experiment, 
we observe a ``perfectly" 
translating finger for the duration of observation. This makes (3.\ref{2.1}) again too 
sharp a question, and the 
theory might well allow for $f$'s similar to those of (3.\ref{2.25}) for arbitrary 
$\lambda$, but decorated by dynamical corrections of $O(e^{-L_0})$ over the finger regime.

In a way, such a possibility is curious. It implies that for (3.\ref{2.25}) a powerful 
surface tension is requested on 
the efflux, while (3.\ref{2.25}) decorated by $O(e^{-L_0})$ corrections produces 
exponentially close motion throughout the 
great bulk of the fluid, while requiring no surface tension at all. This is 
surprising since by (3.\ref{2.30}), for any $\lambda$,
$$\hat{\xi}\sim L_0 -(\frac{1}{\lambda}-1)\varphi,$$
so that macroscopically different fluxes are elicited without physical agency.

Regrettably, beyond our $\lambda=1/2$ exact solutions, we know of no way to construct 
{\em any} other exact solutions with 
two interfaces. For example, it is not difficult to show that the pole dynamics 
of the literature is exhausted by just (3.\ref{2.42}). (Since the details of
this excersize are tangential
to our main thrust here, it has been relegated to an appendix, Section 7.)
This leaves us in the unfortunate position of a merely perturbative 
treatment, which we now erect.

Perturbing a solution to $(2.14')$ determines a very special change of variables 
which renders the calculations 
feasible. Donate by $\hat{f}$ a known solution to $(2.14')$, call its perturbed value $f_p$, 
the strength of perturbation $\epsilon$, and $f$ the perturbation:
\be
f_p=\hat{f}+\epsilon f + ... \label{3.1}
\ee
We shall denote by ``transpose" the symmetry operation of $(2.14')$:
\be
f^t (\zeta,\varphi)\equiv f(-\zeta,\varphi), \label{3.2}
\ee
so that $(2.14')$ for $\hat{f}$ is
\be
\hat{f}'^t \hat{f}_\varphi +(\hat{f}'^t \hat{f}_\varphi)^t =2. \label{3.3}
\ee
Entering (\ref{3.1}) into $(2.14')$ leads to $O(\epsilon)$:
$$\hat{f}'^t f_\varphi +\hat{f}_\varphi^t f'+{\rm transpose}=0,$$
or,
\be
\hat{f}'^t f_\varphi +\hat{f}_\varphi^t f'=\tilde{a}(\zeta,\varphi)\; ; \;
\tilde{a}^t =-\tilde{a}. \label{3.4}
\ee
Integrating (\ref{3.4}) on curves in a solution surface, with $s$ a parameter,
\be
\frac{d\varphi}{ds}=\hat{f}'(-\zeta,\varphi),\frac{d\zeta}{ds}=\hat{f}_\varphi
(-\zeta,\varphi),\frac{df}{ds}=\tilde{a}(\zeta,\varphi). \label{3.5}
\ee
By the first two,
\be
0=\hat{f}_\varphi (-\zeta,\varphi)\frac{d\varphi}{ds}-\hat{f}'(-\zeta,\varphi)
\frac{d\zeta}{ds}=\frac{d}{ds}\hat{f}(-\zeta,\varphi). \label{3.6}
\ee
That is, we should take a function purely of $\hat{f}^t$ as one new variable, and take a 
function of $(\zeta,\varphi)$ independent of $\hat{f}^t$ as $s$. 
Then, the partial derivative of $f$ with respect to this second new 
variable is, by the third o.d.e., 
antisymmetric in $\zeta$. This optimally selects the transpose of the first 
variable as the second variable. To 
ward off exponentials when we consider an $\hat{f}$ such as (3.\ref{2.22}), we are led to 
define the following two new variables in place of $(\zeta,\varphi)$:
\be
\nu \equiv e^{\hat{f}(-\zeta,\varphi)}\; ,\; \xi \equiv e^{\hat{f}(\zeta,\varphi)}. 
\label{3.7}
\ee
We then have by (\ref{3.2}) for any function $u$,
\be
u^t(\nu,\xi)=u(\xi,\nu), \label{3.8}
\ee
the reason we have denoted the symmetry of $(2.14')$ as transpose. The last of 
(\ref{3.5}) with $s\equiv \ln\xi$ is
$$\xi f_2(\nu,\xi)=\tilde{a}(\nu,\xi)\equiv \xi\nu a_{12}(\nu,\xi),\; a^t=-a$$
since the mixed derivative of an antisymmetric function is also antisymmetric, 
and so
\be
f=\nu\partial_\nu a(\nu,\xi). \label{3.9}
\ee
(The integration constant, a function of $\nu$ is just an 
$a(\nu,\xi)=u(\nu)-u(\xi)$.) That is, the 
solution to first order perturbation, (\ref{3.4}) 
are $\ln\nu$ derivatives of an arbitrary antisymmetric ``potential" $a$, and first 
order theory for one interface has 
been fully integrated, {\em despite} the non-local (in $\zeta$) form of the p.d.e. 
$(2.14')$.

Let us check that our new variables $\nu ,\xi$ are sensibly defined. Now, $\zeta$ is 
complex while $\varphi$ is real. Both $\nu$ and 
$\xi$ are complex. If they are formally independent, then upon inverting they determine 
a complex extension of $\varphi$. 
But they are independent, precisely in virtue of $(2.14')$:
\be
\frac{\partial(\ln\xi,\ln\nu)}{\partial(\zeta,\varphi)}=
\left(\begin{array}{c}
 \; \hat{f}'  \; \; \; \hat{f}_\varphi \\
-\hat{f}'^t \; \;  \hat{f}_\varphi^t
\end{array}\right) \label{3.10}
\ee
and so, by $(2.14')$
\be
\left|
\frac{\partial(\ln\xi,\ln\nu)}{\partial(\zeta,\varphi)}
\right|=2 . \label{3.11}
\ee
Inverting (\ref{3.10}),
\be
\frac{\partial(\zeta,\varphi)}{\partial(\ln\xi,\ln\nu)}=\frac{1}{2}
\left(\begin{array}{c}
\hat{f}_\varphi^t  -\hat{f}_\varphi \\
\hat{f}'^t \; \; \; \hat{f}'
\end{array}\right). \label{3.12}
\ee

Paying attention to (3.\ref{2.23}), rather than $\varphi$, $e^{-2\varphi}$ will constantly 
appear, and so we choose for later a new time variable
\be
\lambda\equiv e^{-2\varphi}, \label{3.13}
\ee
for which we have by the second row of (\ref{3.12}),
\be
\nu\partial_\nu \lambda=-\lambda \hat{f}' \; {\rm and}\;
\xi\partial_\xi \lambda=-\lambda \hat{f}'^t \label{3.14}
\ee
which will shortly become important.

Before proceeding it is useful to rewrite $(2.14')$ in $(\nu,\xi)$ variables. By 
(\ref{3.10}),
\be
\partial_\zeta=\hat{f}'\xi\partial_\xi-\hat{f}'^t \nu\partial_\nu,\;
\partial_\varphi=\hat{f}_\varphi \xi\partial_\xi+\hat{f}_\varphi^t \nu\partial_\nu.
\label{3.15}
\ee
Substituting in $(2.14')$ for $\hat{f}$, then yields
\be
\frac{1}{\nu\xi}=f_\xi f_\xi^t-f_\nu f_\nu^t. \label{3.16}
\ee

Clearly $\hat{f}=\ln\xi$ obeys (\ref{3.16}), while $i\ln\nu$ is not reflection symmetric. 
It is now trivial to obtain the perturbation results, which to second order are
\be
f_p=\ln\xi+\epsilon f +\epsilon^2 F+ ... \label{3.17}
\ee
and so,
\be
\frac{1}{\nu}f_\xi+\frac{1}{\xi}f_\xi^t=0 \label{3.18}
\ee
and
\be
\frac{1}{\nu}F_\xi+\frac{1}{\xi}F_\xi^t= f_\nu f_\nu^t -f_\xi f_\xi^t . \label{3.19}
\ee
(\ref{3.18}) is, of course, the observations below (\ref{3.6}), and its general solution 
just (\ref{3.9}). It is also easy, by 
successive integrations by parts and the use of (\ref{3.18}) to integrate 
(\ref{3.19}) fully into
\be
F=f\xi\partial_\xi f+\nu\partial_\nu (\frac{1}{2}f f^t +A) \label{3.20}
\ee
with $A$ another antisymmetric potential.

So far we have just discussed one interface. This suffices for the infinite 
channel. With the flat solution, (3.\ref{2.13}), 
$$\xi=e^{\zeta+\varphi},\; \nu=e^{-\zeta+\varphi}$$
and so
$$a=\frac{\nu^p}{p}-\frac{\xi^p}{p}\;\Rightarrow\;f=\nu^p=e^{-p\zeta+p\varphi}$$
with $p$ the positive integers is a full basis of perturbations for the one interface 
at Re$\zeta=0$, and all are exponentially unstable.

For the 1/2 finger solution of (3.\ref{2.22}) (and almost identically for (3.\ref{2.42})), 
with $\lambda$ of (\ref{3.13}),
$$\xi=\frac{1+e^{\zeta}}{\lambda}, \; \nu=\frac{1+e^{-\zeta}}{\lambda}$$
and again
$$a=\frac{\nu^p}{p}-\frac{\xi^p}{p}\;\Rightarrow\;f=\nu^p=(1+e^{-\zeta})^p e^{2p\varphi}$$
all with $p>0$ unstable. Here we can also take for the $\nu$ term of $a$
$\int\frac{d\nu}{\nu}\ln\nu$ producing $$f=\ln\nu=2\varphi+\ln (1+e^{-\zeta})$$
and so perturb to the nearby Saffman-Taylor solutions (3.\ref{2.25}) with
$\lambda=(1-\epsilon)/2$.  Indeed, any function $u(\nu)$ analytic at 
Re$\zeta\rightarrow +\infty$ provides a legal perturbation.

Let us now go on to the second interface.  Not only does $\hat{f}$ satisfy $(2.14')$,
but so too does
\be
\hat{g}(\zeta,\varphi)=\hat{f}(\zeta+\hat{\xi}(\varphi),\varphi). \label{3.21}
\ee
Accordingly, we define
\be
\tilde{\nu} \equiv e^{\hat{g}(-\zeta,\varphi)}\; ,\; 
\tilde{\xi} \equiv e^{\hat{g}(\zeta,\varphi)}. 
\label{3.22}
\ee
so that, by (\ref{3.21}),
\be
\tilde{\xi}(\zeta,\varphi)=\xi(\zeta+\hat{\xi},\varphi)\; {\rm and}\;
\tilde{\nu}(\zeta,\varphi)=\xi(-\zeta+\hat{\xi},\varphi). \label{3.23}
\ee
It now follows that
\be
g_p=\ln\tilde{\xi}+\epsilon g +\epsilon^2 G+ ... \label{3.24}
\ee
and so by (\ref{3.9})
\be
g=\tilde{\nu}\partial_{\tilde{\nu}} a_g(\tilde{\nu},\tilde{\xi}), \label{3.25}
\ee
and by (\ref{3.20})
\be
G=g\tilde{\xi}\partial_{\tilde{\xi}} g+\tilde{\nu}\partial_{\tilde{\nu}}
 (\frac{1}{2}g g^t +A_g)
 \label{3.26}
\ee
with $a_g$ and $A_g$ antisymmetric in $(\tilde{\nu},\tilde{\xi})$.

But $f_p$ and $g_p$ are related by a $\zeta$ translation, by (2.\ref{15})
\[
\hspace{20mm}
g_p(\zeta,\varphi)=f_p(\zeta+\xi_g(\varphi),\varphi) \hfill (2.\ref{15})
\]
Here we must pause. Should $\xi_g$ be just $\hat{\xi}$ that relates $\hat{f}$ to $\hat{g}$, 
or should it be modified, basically in 
proportion to the strength of the perturbation $f$? Should $\hat{\xi}$ be modified to
\be
\xi_g=\hat{\xi}+\epsilon\Psi+\epsilon^2 \Phi+... \label{3.27}
\ee
(each term a function of $\varphi$),
then certainly these fluctuations can not be freely imposed, since under a fixed 
pressure difference, $p_g$, 
changing $\hat{\xi}$ changes the flux of the flow, and so costs power of an amount 
proportional to the size of the 
fluctuation, and makes it possible for the external agencies driving the flow to 
control the fluctuations. 
Should $\Psi$, etc., however, vanish then the corresponding fluctuation is free and 
uncontrollable.

To be general, substitute (\ref{3.27}) into (2.\ref{15}) and expand in $\epsilon$ to relate $g$ 
and $G$ to $f$ and $F$:
\bee
g(\zeta,\varphi)&=&(f+\hat{f}'\Psi)(\zeta+\hat{\xi},\varphi) \label{3.28} \\
G(\zeta,\varphi)&=&(F+\hat{f}'\Phi+\Psi\partial_\zeta(f+\frac{1}{2}\hat{f}'\Psi))
(\zeta+\hat{\xi},\varphi) \label{3.29} 
\eee
where $\partial_\zeta \Psi=0$ since $\Psi$ is a function just of $\varphi$.

Start with (\ref{3.28}):
$$g(\zeta-\hat{\xi},\varphi)=f+\hat{f}'\Psi.$$
Let us define for convenience
\be
\Psi\equiv\lambda\psi'(\lambda). \label{3.30}
\ee
Then, by (\ref{3.14})
\be
\hat{f}'\Psi=-\nu\partial_\nu \lambda\psi '(\lambda)=-\nu\partial_\nu \psi(\lambda).
\label{3.31}
\ee
But then, by (\ref{3.9}),
\be
\nu\partial_\nu (a-\psi(\lambda))=g(\zeta-\hat{\xi},\varphi). \label{3.32}
\ee
This scheme succeeds because of an extraordinary property of the $(\nu,\xi)$ 
variables.

By (\ref{3.23}),
\bee
\xi_*&=&\tilde{\xi}(\zeta-\hat{\xi},\varphi)=\xi(\zeta,\varphi)=\xi \;{\rm and} \nonumber \\
\nu_*&=&\tilde{\nu}(\zeta-\hat{\xi},\varphi)=\xi(2\hat{\xi}-\zeta,\varphi). \label{3.33}
\eee
So, by (\ref{3.11}), valid just as well for $(\tilde{\xi},\tilde{\nu})$,
$$\left|\frac{\partial(\ln\xi_*,\ln\nu_*)}{\partial(\zeta,\varphi)}\right|=
\left|\frac{\partial(\ln\tilde{\xi},\ln\tilde{\nu})}{\partial(\zeta,\varphi)}\right|_
{(\zeta-\hat{\xi},\varphi)}\left|\begin{array}{c} 1 -\hat{\xi}' \\ 0 \;\;\;\; 1 \end{array}
\right|=2 .$$ 
But then, dividing by (\ref{3.11})
$$\left|\frac{\partial(\ln\xi_*,\ln\nu_*)}{\partial(\ln\xi,\ln\nu)}\right|=1=
\left|\frac{\partial(\ln\xi,\ln\nu_*)}{\partial(\ln\xi,\ln\nu)}\right|=
\frac{\partial\ln\nu_*}{\partial\ln\nu}$$
and so,
\be
\nu_*=\xi(2\hat{\xi}-\zeta,\varphi)=\nu/\gamma(\xi). \label{3.34}
\ee
The remarkable result is that $\gamma$ is a function purely of $\xi$.

Let us elaborate on (\ref{3.34}). The simplicity of its deduction reflects the curious fact 
that the equations of motion 
are just the constant Jacobian of $(f(\zeta,\varphi),f(-\zeta,\varphi))$ with respect to 
$(\zeta,\varphi)$. It is straightforward 
to directly deduce it by the joint integration of $(2.14')$ for both $f(\zeta,\varphi)$ and 
$g(\zeta,\varphi)=f(\zeta+\xi_g,\varphi)$ . 

Writing
\be
\gamma(\xi)=e^{\Gamma(\ln\xi)}=e^{\Gamma(f(\zeta,\varphi))}, \label{3.34'}
\ee
assuming (\ref{3.34}) we have
\be
f(-\zeta,\varphi)-f(2\xi_g-\zeta,\varphi)=\Gamma(f(\zeta,\varphi)) \label{3.34"}
\ee
where again, it is remarkable that $\Gamma$ is a function purely of $f(\zeta,\varphi)$. 

To directly deduce this result, write down 
$(2.14')$ for $g=f(\zeta+\xi_g)$ and translate the argument $\zeta\rightarrow\zeta-\xi_g$. 
Subtract from this $(2.14')$ for $f$ and so obtain
$$0=f'(\zeta)\partial_\varphi(f(-\zeta)-f(2\xi_g-\zeta))$$
$$-f_\varphi(\zeta)\partial_\zeta(f(-\zeta)-f(2\xi_g-\zeta)).$$
Dividing by $f'f_\varphi$, the equality of ratios is equivalent to (\ref{3.34"}). That is, 
if just $f$ obeys $(2.14')$ {\em and} 
satisfies (\ref{3.34"}), then $f$ is a solution for the finite problem, and 
conversely. This will enable us to 
put severe restrictions on the nature of solutions near to $\hat{f}$, a pure 1/2 
finger solution; sufficiently severe to prove pattern selection.

Now, by (\ref{3.25}), 
$$g(\zeta,\varphi)=\tilde{\nu}a_{g,1}(\tilde{\nu},\tilde{\xi})$$ 
and so,
\bee
g(\zeta-\hat{\xi},\varphi) &=& \nu_* a_{g,1}(\nu_*,\xi_*)=\frac{\nu}{\gamma(\xi)}a_{g,1}
(\frac{\nu}{\gamma(\xi)},\xi) \nonumber \\
&=& \nu\partial_\nu a_g(\frac{\nu}{\gamma(\xi)},\xi). \label{3.35}
\eee
But now, by (\ref{3.32}),
\be
a(\nu,\xi)-\psi(\lambda)-\rho(\xi)=a_g(\frac{\nu}{\gamma(\xi)},\xi) \label{3.36}
\ee
with the integration constant $\rho$ purely a function of $\xi$, and we have 
succeeded in simultaneous integrating both equations of motion.

At this point a few observations need be made. First $\lambda$ is independent of $\zeta$. 
Writing $\lambda(\nu,\xi)$, the value is 
unchanged by sending $\zeta\rightarrow -\zeta$, and so $\lambda$ is symmetric in $\nu$ and 
$\xi$:
\be
\lambda(\nu,\xi)=\lambda(\xi,\nu). \label{3.37}
\ee
(This is implicit, of course, in (\ref{3.14}).)  Next, write (\ref{3.34}) as
\be
\gamma(\xi(\zeta,\varphi))=\frac{\xi(-\zeta,\varphi)}{\xi(2\hat{\xi}-\zeta,\varphi)}.
\label{3.38}
\ee
But then
$$\gamma(\nu_*)=\gamma(\frac{\nu}{\gamma(\xi)})=\gamma(\xi(2\hat{\xi}-\zeta))=
\frac{\xi(\zeta-2\hat{\xi})}{\xi}=\frac{\nu(2\hat{\xi}-\zeta)}{\xi}$$
and so, together with (\ref{3.34}), we have 
\be
\nu(2\hat{\xi}-\zeta,\varphi)=\xi\gamma(\frac{\nu}{\gamma(\xi)}),\;
\xi(2\hat{\xi}-\zeta,\varphi)=\frac{\nu}{\gamma(\xi)}. \label{3.39}
\ee
But then with $\lambda$ independent of $\zeta$,
\be
\lambda(\nu,\xi)=\lambda(\xi\gamma(\frac{\nu}{\gamma(\xi)}),\frac{\nu}{\gamma(\xi)}).
\label{3.40}
\ee
Substituting $\nu\rightarrow\nu\gamma(\xi)$,
\be
\lambda(\nu\gamma(\xi),\xi)=\lambda(\xi\gamma(\nu),\nu)\equiv\Lambda(\nu,\xi):\;
\Lambda^t=\Lambda \label{3.41}
\ee
and $\Lambda$ is another symmetric function of $\nu$ and $\xi$. Substituting 
$\nu\rightarrow\nu\gamma(\xi)$ in (\ref{3.36}) now yields
\be
a(\nu\gamma(\xi),\xi)-\psi(\Lambda)-\rho(\xi)=a_g(\nu,\xi).  \label{3.42}
\ee
Transposing (\ref{3.42}) and adding, by the antisymmetry of $a_g$ we have
\be
a(\nu\gamma(\xi),\xi)+a(\xi\gamma(\nu),\nu)-2\psi(\Lambda)-\rho(\xi)-\rho(\nu)=0,
\label{3.43}
\ee
and we have determined what $a$'s in (\ref{3.9}) are allowed for the problem with two 
interfaces.

Let us see that the decomposition of (\ref{3.36}) is unique, so that for a given 
perturbation $f$, and hence a given 
function $a$, both $\psi(\lambda)$ and $\rho(\xi)$ are uniquely determined. 
(Were this not the case then $\Psi$ would lose meaning.) 
With $\Lambda$ and $\rho$ denoting the differences of two different decompositions, we 
have
\be
-2\psi(\Lambda)=\rho(\xi))+\rho(\nu)\; (i.e. \; a=0). \label{3.44}
\ee
But, with $a=0$, (\ref{3.36}) is
$$a_g(\nu/\gamma,\xi)=-\psi(\lambda)-\rho(\xi),$$
and by (\ref{3.35}), (\ref{3.31})
$$g(\zeta-\hat{\xi})=-\nu\partial_\nu \psi=\hat{f}'\Psi.$$
But then
$$g_p(\zeta-\hat{\xi})=\hat{f}+\epsilon\hat{f}'\Psi=\hat{f}(\zeta+\epsilon\Psi),$$
or
\be
g_p=\hat{f}(\zeta+(\hat{\xi}+\epsilon\Psi),\varphi)\; {\rm and}\; f_p=\hat{f} \;
(a=f=0). \label{3.45}
\ee
That is, the $\psi$ of (\ref{3.44}) corresponds to the identical unperturbed motion $\hat{f}$ 
throughout the fluid, but we 
have taken a new interface at $\hat{\xi}_p=\hat{\xi}+\epsilon\Psi$. This means that we have 
simply considered a larger finite body of fluid, 
and so $\psi$ of (\ref{3.44}) is the perturbation of changing the volume of fluid. This is 
not what we consider, since 
the physical experiment is predicated on volume preservation. That is, for the 
physical perturbations we 
care about, which contain no part of $\rho$ of the form of that which is 
determined by (\ref{3.44}), our decomposition is unique.

At the end of Section 3, in (3.\ref{2.13}), (3.\ref{2.22}) and (3.\ref{2.42}) we recorded 
the fluid volume through the parameter $k$, 
which is $e^{2L_0}$ with $2\pi L_0$ the finite area of fluid. In changing $L_0$ by 
$\epsilon$, we change $k$ by $2\epsilon k$ and so the $\Psi$ in (\ref{3.45}) is
\be
\Psi_L=k\frac{\partial(2\hat{\xi})}{\partial k}. \label{3.46}
\ee
Thus we have for the flat solution by (3.\ref{2.13})
\be
\Psi_L=1,\; \psi_L=\ln\lambda \label{3.47}
\ee
and for both the $\lambda=1/2$ fingers of (3.19) and (3.\ref{2.42})
\be
\Psi_L=\frac{k\lambda}{1+k\lambda},\; \psi_L=\ln(1+k\lambda). \label{3.48}
\ee

Let us now put in evidence the functions $\lambda,\; \gamma\; {\rm and}\; \Lambda$ 
for these same solutions.

For (3.\ref{2.13}),
\bee
\lambda &=& 1/\xi\nu, \nonumber \\
\gamma &=& e^{-2\hat{\xi}}=1/k  \label{3.49} \\
{\rm and} \;\; \Lambda &=& k/\xi\nu. \nonumber
\eee

Similarly, for (3.\ref{2.22}), 
\bee
\lambda &=& \frac{1}{\xi}+\frac{1}{\nu}, \nonumber \\
\gamma &=& \frac{\xi}{\xi+k} \label{3.50} \\
{\rm and} \;\; \Lambda &=& \frac{1}{\xi}+\frac{1}{\nu}+\frac{k}{\xi\nu} . \nonumber
\eee

Finally, for dynamical (3.\ref{2.42}), 
\bee
\lambda &=& \frac{1}{\xi}+\frac{1}{\nu}+\frac{1}{\xi\nu} , \nonumber \\
\gamma &=& \frac{\xi+1}{\xi+k} \label{3.51} \\
{\rm and} \;\; \Lambda &=& \frac{1}{\xi}+\frac{1}{\nu}+\frac{k}{\xi\nu} . \nonumber
\eee

Now, by (\ref{3.44}) and (\ref{3.47}), (\ref{3.48}) we can determine the forbidden forms of 
$\rho$ that generate pure volume changes.
For (3.\ref{2.13})
\be
\rho_L(\xi)=2\ln\xi-\ln k, \label{3.52}
\ee
and for both (3.\ref{2.22}) and (3.\ref{2.42}), 
\be
\rho_L(\xi)=-2\ln(1+k/\xi). \label{3.53}
\ee

Although we have put these details in explicit evidence for the exact solutions 
we know of, it is to be 
stressed that the entire machinery of this section is directly applicable to the 
perturbation on {\em any} $\hat{f},\hat{g}$ 
which solve $(2.14')$. In particular, it will apply to the perturbations of our 
explicit solutions, now regarded 
as new fundamental solutions, up to the order we consider. 

We now need to determine $a(\nu,\xi)$ from (\ref{3.36}), or eliminating $a_g$, 
from (\ref{3.43}). With $a$ antisymmetric, we can 
always write
\be
a(\nu,\xi)\equiv u(\nu,\lambda)-u(\xi,\lambda). \label{3.54}
\ee
Entering (\ref{3.54}) into (\ref{3.43}) and then substituting 
$\nu\rightarrow\nu/\gamma(\xi)$ and rearranging, we have 
\bee
(u(\nu,\lambda)-u(\frac{\nu}{\gamma(\xi)},\lambda)-\rho(\xi) &-& \psi(\lambda))\nonumber \\
+(u(\xi\gamma(\frac{\nu}{\gamma(\xi)}),\lambda)-u(\xi,\lambda)-
\rho(\frac{\nu}{\gamma(\xi)})&-& \psi(\lambda))=0. \label{3.55}
\eee
This equation is much more transparent than it appears. Reverting to invertible 
$(\zeta,\lambda)$ variables, with 
\be
u(\nu,\lambda)=u(\xi(-\zeta,\lambda),\lambda)\equiv h(\zeta,\lambda), \label{3.56}
\ee
(\ref{3.54}) is
\be
a(\zeta,\lambda)=h(\zeta,\lambda)-h(-\zeta,\lambda) \label{3.54'}
\ee
and (\ref{3.55}), by (\ref{3.39}) is
\[
(h(\zeta,\lambda)-h(\zeta-2\hat{\xi},\lambda)-\rho(\xi(\zeta,\lambda))-\psi(\lambda))+
\]
\[
(h(2\hat{\xi}-\zeta,\lambda)-h(-\zeta,\lambda)-\rho(\xi(2\hat{\xi}-\zeta,\lambda))
-\psi(\lambda))=0
\] 
or, with the definition of $r(\zeta,\lambda)$:
\be
h(\zeta,\lambda)-h(\zeta-2\hat{\xi},\lambda)\equiv\rho(\xi(\zeta,\lambda))
+\psi(\lambda)+r(\zeta,\lambda), \label{3.57}
\ee
\be
r(\zeta,\lambda)+r(2\hat{\xi}-\zeta,\lambda)=0. \label{3.55'}
\ee
It is now easy to see that $r$ can be taken to vanish with impunity. This 
follows because we are uninterested 
in $h$ itself, but rather by (\ref{3.54'}), its antisymmetric part.  But, (\ref{3.57}) is
linear in $h$, and so can be decomposed into three pieces of $h$:  One piece satisfying
(\ref{3.57}) with $r\equiv 0$, a second with $\rho+\psi\equiv 0$ with $r$ satisfying 
(\ref{3.55'}), and finally a purely homogeneous part
$$h_0(\zeta,\lambda)=h_0(\zeta-2\hat{\xi},\lambda)$$
and so with
\be
a_0(\zeta,\lambda)=a_0(\zeta-2\hat{\xi},\lambda). \label{3.58}
\ee
As we have already discussed, an antisymmetric doubly periodic function must have 
singularities within 
physical fluid, so that $a_0$ can be dropped by simply determining the rest of  
$h$ to have correct analyticity. 
Returning to the second $r$ contributed piece of $h$, $h_r$, and suppressing the 
$\lambda$-dependence, 
$$h_r(\zeta)-h_r(\zeta-2\hat{\xi})=r(\zeta).$$
Setting $\zeta\rightarrow 2\hat{\xi}-\zeta$, and using (\ref{3.55'}),
$$h_r(-\zeta)-h_r(2\hat{\xi}-\zeta)=r(\zeta)=h_r(\zeta)-h_r(\zeta-2\hat{\xi}),$$
or 
$$a_r(\zeta,\lambda)=a_r(\zeta-2\hat{\xi},\lambda)$$
which then can be totally absorbed into $a_0$, and then $a_0$ dropped completely for an $h$ 
with correct analyticity, which then simply obeys
$$h(\zeta,\lambda)-h(\zeta-2\hat{\xi},\lambda)=\rho(\xi(\zeta,\lambda))+\psi(\lambda).$$

For a convenience of geometry (what we refer to mentally as Re$\zeta\rightarrow +\infty$)
, it is 
convenient to redefine $h(\zeta,\lambda)\rightarrow -h(-\zeta,\lambda)$, which 
leaves (\ref{3.54}) and (\ref{3.54'}) unchanged, and produces, after sending 
$\zeta\rightarrow -\zeta$
\be
h(\zeta+2\hat{\xi},\lambda)-h(\zeta,\lambda)=\rho(\nu(\zeta,\lambda))+\psi(\lambda).
\label{3.59}
\ee
The unique solution of (\ref{3.59}) - the one which possesses correct annular 
analyticity - is then the unique solution to (\ref{3.43}).

Let us next note that for the exact solutions of (\ref{3.49})-(\ref{3.51}), 
as Re$\zeta\rightarrow +\infty$, $\nu$ vanishes for the flat solution and has, 
for the other two 1/2 finger solutions
\be
\nu\rightarrow 1/\lambda \; {\rm as \; Re}\zeta\rightarrow +\infty, \label{3.60}
\ee
so that solutions to (\ref{3.59}) with $h$ regular as Re$\zeta\rightarrow +\infty$, 
\be
\psi(\lambda)=-\rho(1/\lambda), \label{3.61}
\ee
for the 1/2 finger solutions, but
\be
\psi(\lambda)=-\rho(0) \;\Rightarrow\; \Psi=0 \label{3.62}
\ee
for the flat solution. For the analogs of the infinite fluid problem, with 
$\rho=\nu^n/n,\; \Psi=\lambda\psi '(\lambda)=e^{2n\varphi}$, 
and so nothing is changed 
for the flat solution's perturbations, but now for perturbations about the 1/2 
finger, impedance (flux) grows 
exponentially in proportion to the instability's strength, and so are not free, 
and controllable by the pump 
that drives the flow: It is exactly as potent to control the change of flux as 
to control the growth of these 
instabilities. This is totally different from the infinite fluid case where 
these instabilities were freely 
impressible and uncontrollable. At this point we already mathematically conclude 
that the geometry of 
infinite fluid is physically incorrect, and the limit of large length clearly 
singular. This is the first main 
point we set out to establish.

Let me emphasize this. {\em The literature has been calculated in a physically 
incorrect geometry, and in 
consequence, all its conclusions lie suspect.} (Its incorrectness in grander ways 
will transpire.)

As for $h$'s that are regular as Re$\zeta\rightarrow -\infty$, all $\nu$'s 
of the exact solutions diverge, and so the $\rho$'s that determine 
them have constant limits (for simple powers, zero), and so these all have $\Psi=0$. 
These will all turn out, as do 
the powers which relax as $e^{-2n\varphi}$, as relaxing stabilities. 
Thus, the flat solution is uncontrollably unstable, but 
the 1/2 finger solution can surely, under appropriate experimental control, be 
reduced to non-exponential 
slow modes. It is our goal to show that these putative slow modes are 
non-existent.

At this point we drop any further focus upon the flat solution, and restrict 
ourselves to the dynamical 1/2 
finger of (\ref{3.51}), since (\ref{3.50}) is simply a long-time, long-fluid limit of this. 
We will now write down all 
possible ``perturbations" $f$ upon this solution.

Denote by $x, y$:
\be
y\equiv e^{\zeta-\hat{\zeta}(\lambda)},\; x\equiv y(-\zeta,\lambda)\Rightarrow
xy=e^{-2\hat{\zeta}}=1+\lambda \label{3.63}
\ee
so that
\be
\nu=(1+x)/\lambda,\;\xi=(1+y)/\lambda . \label{3.64}
\ee
Denote by $\beta(\lambda)$:
\be
\beta(\lambda)\equiv e^{-2\hat{\xi}}=\frac{1+\lambda}{1+k\lambda} \label{3.65}
\ee
by (3.\ref{2.42}). Finally, define
\be
v(x,\lambda)\equiv h(\zeta,\lambda), \label{3.66}
\ee
so that (\ref{3.59}) reads
\be
v(\beta x,\lambda)-v(x,\lambda)=\rho(\frac{1+x}{\lambda})+\psi(\lambda), \label{3.67}
\ee
and 
\be
a(\nu,\xi)=v(x,\lambda)-v(y,\lambda). \label{3.54"}
\ee
By straightforward differentiation, suppressing the $\lambda$-dependence,
$$f=(1+x)v'(x)-$$
\be
\frac{y}{1+y}((1+x)v'(x)+\lambda v_\lambda(x)-(x\rightarrow y)). 
\label{3.68}
\ee
The flux-bearing solutions, $x\rightarrow 0$, are, by summing (\ref{3.67}) at the 
contractive ($\beta<1$) fixed point at 0,
\bee
v_d(x,\lambda)=-\sum_0 (\rho(\frac{1+\beta^n x}{\lambda})-\rho(\frac{1}{\lambda})); 
\label{3.69}
\\ \psi=-\rho(\frac{1}{\lambda})\Rightarrow \Psi=\frac{1}{\lambda}\rho'(\frac{1}{\lambda})
\nonumber
\eee
which we call ``down-summing", or ``$d\Sigma$".

The flux-free stabilities, $x\rightarrow \infty$, are, by summing the inverse,
\be
v_u(x,\lambda)=-\sum_1 (\rho(\frac{1+\beta^{-n}x}{\lambda})-\rho(\infty)),\; \Psi=0 
\label{3.70}
\ee
which we call ``up-summing", or ``$u\Sigma$". (It is immediate to verify that 
the $\zeta$ derivative of $v_d+v_u$ is precisely 
doubly-periodic when both sums exist.) Should $\rho$ be neither up nor down 
summable, then we simply split 
it into $\rho=\rho_u+\rho_d-\psi_\rho(\lambda)$ with $\psi_\rho$, a piece of the 
$\psi$ for this $\rho$. 
(This is just a Laurent expansion when not explicitly 
possible.) Thus, we can determine all possible $v$'s.

Let us now write
\be
R(\nu)\equiv -\nu\rho '(\nu). \label{3.71}
\ee
For $d\Sigma$ solutions, by (\ref{3.69}),
\be
\Psi(\lambda)=-R(1/\lambda). \label{3.72}
\ee
To determine $f$, we need one last object, which by (\ref{3.65}) is
\be
-\lambda\frac{\beta '}{\beta}\equiv \alpha(\lambda)=\frac{1-\beta}{1+\lambda}. \label{3.73}
\ee
Then, by (\ref{3.68}),
\bee
f_d=R(\nu)+\sum_1 \frac{(1+x)\beta^n}{1+\beta^n x}R(\frac{1+\beta^n x}{\lambda})+ 
\nonumber \\
\frac{y}{1+y}\sum_1 (\frac{1-\beta^n+n x \beta^n \alpha(\lambda)}{1+\beta^n x}
R(\frac{1+\beta^n x}{\lambda})-R(\frac{1}{\lambda})) \nonumber \\
-\frac{y}{1+y}\sum_1 (x\rightarrow y). \label{3.74}
\eee

This equation, in conjunction with (\ref{3.72}) is remarkable: It says that the pump's 
measurement of extra 
requested flux as a function of time uniquely determines and is uniquely 
determined by the {\em entire} spatial 
shape of the flow: The pump can literally be programmed to create or 
reciprocally control any desired unstable conformation of the fluid!

Next, in the finger regime $\beta\sim 1/k\lambda$ and is astronomically, exponentially 
small, so that the sums in (\ref{3.74}) are 
astronomically small, so that $f_d\sim R(\nu)$, just the result of the infinite fluid with 
one interface. For example, $R=\ln(1+a\nu)$ produces a pole-dynamics singularity
\be
R=\ln(1+\frac{a}{\lambda}(1+e^{-\zeta-\hat{\zeta}}))=\ln(1+\frac{a}{\lambda})+
\ln(1-e^{\zeta_s-\zeta}) \label{3.75}
\ee
with 
$$e^{\zeta_s}=\frac{-e^{-\hat{\zeta}}}{1+\lambda/a} $$
which for $a\in (0,2)$ is always $\zeta_s<0$ for all $\lambda$, and for $a=1$ 
is just $\zeta_s=\hat{\zeta}$. Clearly, as $\beta\rightarrow 0$ for a long enough channel 
during the finger regime, we have then a propagating finger of width 
$(1-\epsilon)/2$, and all fingers are allowed. So it would seem. (It is false.)

In any case, with $R(\nu)$ analytic as Re$\zeta\rightarrow +\infty$, these down sums are 
perturbations exponentially stronger (by roughly $e^{\hat{\zeta}}\sim e^L$) on the 
driven interface than on the efflux interface, and clearly by (\ref{3.72}) all 
instabilities.

Finally, let us notice where all of $f_d$'s singularities are. First, $R(0)=0$ since by 
(\ref{3.71}) this is true unless $\rho=\ln\nu+...$ 
which by (\ref{3.53}) means it contains a piece that changes the area of the fluid, 
which is illegal. Thus the poles 
at $1+\beta^n x=0$ and $1+\beta^n y=0$ are removable. The pole at $y+1=0$, (or $\xi=0$)
is a higher order correction to the basic singularity of $\hat{f}=\ln\xi$, 
and is of no special concern. Since $\beta^n=e^{-2n\hat{\xi}}$, we then see that 
$f_d$ of (\ref{3.74}) has singularities at
\be
\zeta_{+n}=\zeta_s-2n\hat{\xi}\; n=0.. \; {\rm and} \;
\zeta_{-n}=2n\hat{\xi}-\zeta_s\; n=1.. \label{3.76}
\ee
when $R(\nu)$ has a singularity at $\zeta_0=\zeta_s$. Most important is that 
$\zeta_{-0}=-\zeta_s$ is {\em not} a singularity. (By (\ref{3.69}) the 0 term in the sum 
yields $a=\rho(\xi)-\rho(\nu)$ and $f=\nu\partial_\nu a$ 
dispatches the $\rho(\xi)$ term which contains the putative 
singularity at $-\zeta_s$.) In consequence $-\zeta_s$
can lie in physical fluid $(0,\hat{\xi})$, and so, $f_d$'s can have singularities below, 
but arbitrarily close to Re$\zeta=0$, and so able 
to significantly modify the shape of the driven interface. Again, so it would 
seem. 

Also, in particular, $f_d$ is 
singular at $2\hat{\xi}-\zeta_s$ if it is at $\zeta_s$. 
This will prove to be of the utmost consequence.

The up-sum solutions are of a totally different character. By (\ref{3.70}) we have 
\bee
f_u=-\sum_1 \frac{1+x}{\beta^n +x}R(\frac{1+\beta^{-n} x}{\lambda})+ 
\label{3.77} \\
\frac{y}{1+y}\sum_1 \frac{1-\beta^n+n x \alpha(\lambda)}{\beta^n +x}
R(\frac{1+\beta^{-n} x}{\lambda}) \nonumber \\
-\frac{y}{1+y}\sum_1 (x\rightarrow y). \nonumber
\eee
It is useful to note that the sum of the two $n=1$ terms in $x$ produce
\be
f_u=-R(\frac{1+\beta^{-1}x}{\lambda})+..., \label{3.78}
\ee
and $f_u$ in the finger regime is well approximated by just this leading term. As 
for singularities, with $R(\nu)$ possessing a singularity at $\zeta_s$, $f_u$ is 
singular at
\be
\zeta_{\pm n}=\pm (\zeta_s+2n\hat{\xi})\;n=1,2,... \label{3.79}
\ee
However, apart from $\zeta_{+1}$, by (\ref{3.73}) all other terms are also singular at
\be
\zeta_{\pm n}=\pm (-\hat{\zeta}+2n\hat{\xi})\;n=1,2,...\;{\rm but\; not\;}\zeta_{+1}. 
\label{3.80}
\ee
Should $R$ be logarithmic, (i.e. have cuts) such as (\ref{3.75}), then the first term, 
(\ref{3.78}) is
\be
\ln(1-e^{\zeta_s+2\hat{\xi}-\zeta}). \label{3.81}
\ee
This has two immediate consequences. First, $\zeta_s>-\hat{\xi}$ so that 
$\zeta_s+2\hat{\xi}>\hat{\xi}$, and so outside of physical fluid. (For $\zeta_s$ more 
negative than $-2\hat{\xi}$, one of the $\zeta_{+n}$ will still be in fluid.) Thus, such 
perturbations can only be imposed after a time 
with $\lambda$ sufficiently small. Second, with $\zeta_s+2\hat{\xi}>\hat{\xi}$, 
the only branching of (\ref{3.81}) that is reflection symmetric is 
with the cut running off to the left, and so throughout physical fluid, thereby 
violating channel width. 
Accordingly, branched $R(\nu)$'s must contain paired singularities, with cuts beginning 
on one singularity, and terminating on a mating one (or ones). That is, if
$$R=\Sigma \alpha_n \ln(1+a_n \nu),$$
then
\be
\Sigma \alpha=0. \label{3.82}
\ee
With (\ref{3.82}) enforced, and with $R$ logarithms, the long fluid behavior of 
(\ref{3.78}) is that of pole-dynamics 
with singularities to the right of fluid (low pressure) sinks, but sinking 
zero net flux.

For both second order perturbation and the theory of perturbations about 
perturbed solutions, we require 
some facts about first order solutions which follow by differentiating (\ref{3.41}) 
and (\ref{3.43}). Differentiating (\ref{3.41}) by $\xi\partial_\xi$, and
setting $\nu\rightarrow\nu/\gamma$ yields by (\ref{3.14}) and (\ref{3.39})
\be
\frac{\xi\gamma '}{\gamma}\hat{f}'(\zeta)+\hat{f}'(-\zeta)=\hat{f}'(2\hat{\xi}-\zeta),
\label{3.83}
\ee
a result just as well obtained by differentiating $\ln$(\ref{3.38}) on $\zeta$.

Differentiating (\ref{3.43}) by $\xi\partial_\xi$, together with $a$'s antisymmetry 
produces, after noticing that
$$\frac{1}{\Lambda}\xi\partial_\xi\Lambda=\frac{1}{\Lambda}\xi\partial_\xi
\lambda(\xi\gamma(\nu),\nu)=-\hat{f}'(\xi\gamma(\nu),\nu)$$
and setting $\nu\rightarrow\nu/\gamma(\xi)$, 
\bee
\frac{\xi\gamma '}{\gamma}f(\zeta)-f(-\zeta)+f(2\hat{\xi}-\zeta)+2\Psi(\lambda)
\hat{f}'(2\hat{\xi}-\zeta) \nonumber \\
=\xi\rho '(\xi)\equiv -R(\xi).  \label{3.84}
\eee
As a last observation important for second order theory, by (\ref{3.14}),
\be
P\equiv\Psi(\lambda)\hat{f}'=-\nu\partial_\nu\psi\; {\rm and} \;
P^t\equiv\Psi(\lambda)\hat{f}'^t=-\xi\partial_\xi\psi \label{3.85}
\ee
so that 
\be
\xi\partial_\xi P=\nu\partial_\nu P^t. \label{3.86}
\ee
Notice also that 
$$\tilde{\nu}\partial_{\tilde{\nu}}u |_{\zeta-\hat{\xi}}=
\nu\partial_\nu u(\frac{\nu}{\gamma(\xi)},\xi)$$
and
$$\tilde{\xi}\partial_{\tilde{\xi}} u |_{\zeta-\hat{\xi}}=
(\xi\partial_\xi +\frac{\xi\gamma '}{\gamma}\nu\partial_\nu )
u(\frac{\nu}{\gamma(\xi)},\xi).$$

It is now straightforward to write out second order perturbation theory. First 
calculate the pull-back of $G$ of (\ref{3.26}) to $\zeta-\hat{\xi}$. By (\ref{3.28})
$$g(\zeta-\hat{\xi})=g(\frac{\nu}{\gamma},\xi)=(f+P)(\nu,\xi),\; {\rm or}$$ 
$$g(\nu,\xi)=(f+P)(\nu\gamma(\xi),\xi).$$
For $g^t=g(\tilde{\xi},\tilde{\nu})$, we have 
$$g^t(\zeta-\hat{\xi})=g(\xi,\frac{\nu}{\gamma})= 
(f+P)(\xi\gamma(\frac{\nu}{\gamma}),\frac{\nu}{\gamma})=(f+P)(2\hat{\xi}-\zeta).$$
Now, using (\ref{3.84}) and (\ref{3.83}) multiplied by $\Psi$, obtain
$$\frac{1}{2}g g^t(\zeta-\hat{\xi})=$$
$$\frac{1}{2}(f+P)(f^t-P^t)-\frac{\xi\gamma '}{2\gamma}(f+P)^2-\frac{1}{2}(f+P)R(\xi).$$
Combined with the first term of (\ref{3.26}) pulled back to $\zeta-\hat{\xi}$, we have
$$G(\zeta-\hat{\xi})=(f+P)(\xi\partial_\xi f+\nu\partial_\nu P^t)+$$ 
$$\nu\partial_\nu(\frac{1}{2}(f+P)(f^t-P^t-R(\xi))+A_g(\frac{\nu}{\gamma},\xi)).$$
We now equate this expression to $G(\zeta-\hat{\xi})$ as obtained from $f_p$, 
(\ref{3.29}), using (\ref{3.15}) multiplied by $\Psi$. Defining the second order flux
\be
\Phi\equiv\lambda\varphi '(\lambda), \label{3.87}
\ee
after obvious cancellations, we produce an expression of the form $\nu\partial_\nu(\; )$, 
and so, with an integration constant $\mu(\xi)$
purely of $\xi$, obtain the fully integrated second order analog to (\ref{3.36})
\bee
A(\nu,\xi)-\varphi(\lambda)-\mu(\xi)= \label{3.88} \\
\frac{1}{2}f(P^t-R(\xi))+\frac{1}{2}P(f^t-R(\xi))+A_g(\frac{\nu}{\gamma},\xi)).
\nonumber
\eee
We now proceed exactly as we did to achieve (\ref{3.57}), with now an extra right-hand 
driving term. $\varphi$ and $\mu$ 
are chosen as the minimal required terms to produce an $A$ analytic over 
physical fluid. Here we must 
generally avail ourselves of an appropriate doubly periodic homogeneous 
solution, should the particular 
solution determined by the driving term from first order happen to produce 
unacceptable singularities. 
(The form of (\ref{3.88}) with $f^t-R(\xi)$, and the construction of $F$ from 
the potential $A+\frac{1}{2}f f^t$ assists in eliminating such problems.)

To end this section on formalities, let us see how $\gamma$, or $\Gamma$ is modified to 
first order in $\epsilon$ from $\hat{\gamma}$ and $\hat{\Gamma}$. 
Expanding (\ref{3.34"}),
\bee
\hat{f}(-\zeta)-\hat{f}(2\hat{\xi}-\zeta)-2\epsilon\Psi\hat{f}'(2\hat{\xi}-\zeta) 
\nonumber \\
+\epsilon(f(-\zeta)-f(2\hat{\xi}-\zeta)) \nonumber \\
=\Gamma(\hat{f}(\zeta)+\epsilon f(\zeta))+O(\epsilon^2). \label{3.89}
\eee
With $\hat{f}(-\zeta)-\hat{f}(2\hat{\xi}-\zeta)=\hat{\Gamma}(\hat{f}(\zeta))$, 
and dividing by $\epsilon$,
\[
f(-\zeta)-f(2\hat{\xi}-\zeta)=\frac{1}{\epsilon}
(\Gamma(\hat{f}(\zeta)+\epsilon f(\zeta))-\hat{\Gamma}(\hat{f}(\zeta)))
\]
\[
+2\Psi\hat{f}'(2\hat{\xi}-\zeta)+O(\epsilon). 
\]
Since, by (\ref{3.34'}),
\be
\hat{\Gamma} '(\hat{f}(\zeta))=\xi\frac{\gamma '}{\gamma}(\xi), \label{3.90}
\ee
by (\ref{3.84}), the perturbation theory,
\[
\frac{1}{\epsilon}(\Gamma(\hat{f}(\zeta)+\epsilon f(\zeta))-\hat{\Gamma}(\hat{f}(\zeta))) \\
=R(e^{\hat{f}})+\hat{\Gamma}'(\hat{f}(\zeta))f(\zeta)+O(\epsilon).
\]
Taking the limit $\epsilon\sim 0$ ($\Gamma$ must be analytic),
$$\Gamma(\hat{f})=\hat{\Gamma}(\hat{f})+\epsilon R(e^{\hat{f}})+O(\epsilon^2).$$
That is, for $f_p$ to be the full perturbed solution to our problem, and with 
$\xi=e^{f_p(\zeta)}$,
\be
\Gamma(f_p(\zeta))=\hat{\Gamma}(f_p(\zeta))+\epsilon R(\xi)+O(\epsilon^2) \label{3.91}
\ee
or,
\be
\gamma(\xi)=\hat{\gamma}(\xi)(1+\epsilon R(\xi))+O(\epsilon^2). \label{3.91'}
\ee
This result has deep consequences for perturbations, which we now spell out.


\section{Derivatives Versus Perturbations}
\setcounter{equation}{0}

The idea of perturbation theory is that given a solution, we find
all the (partial in $\epsilon$) derivatives of the equations of motion about 
this point
which, in principle, determines a power series for the solution
with leading term the known solution. Should the solution actually be
differentiable over some compact region, then the first order
solution $\hat{f}+\epsilon f$, ($f$ just a particular tangent)
with $\epsilon$ small but finite,
agrees with an actual solution over the region within a uniform
$O(\epsilon^2)$ error. In general one needs either powerful
bounds or the entire series to all orders with the knowledge of its
convergence, to know if this derivative actually exists. It is
only then that $\hat{f}+\epsilon f$ is an
approximation to a solution. And this can almost never be settled
to any finite order in perturbation without exceptional extra
knowledge. The exact result (4.\ref{3.34"}) and it uniquely
determined perturbed form, (4.\ref{3.91}) provides us with some such
knowledge.

Specifically, (4.\ref{3.91'}) says that if $\hat{f}+\epsilon f$ for a
finite, fixed value of $\epsilon$ is an approximation to an actual
solution, then $\gamma$ exists as a function purely of $\xi$, and
the leading term and part of its $O(\epsilon)$ corrections resides
in $\hat{\gamma}(\xi)$. That is, in (3.\ref{3.91}) what must be
true is that the first term is
\be
\hat{\Gamma}(\hat{f}+\epsilon f) \label{3.92}
\ee
and {\em not}
\begin{equation}
\hat{\Gamma}(\hat{f})+\epsilon\hat{\Gamma}'(\hat{f})f+O(\epsilon^2). 
\label{3.92'}
\end{equation}
Evidently, if $f$ is always bounded over the region we care about,
then (\ref{3.92'}) is altogether equivalent to (\ref{3.92}). Should $f$
not be bounded over the region of control then only
(\ref{3.92}) applies, and (\ref{3.92'}) is wrong. The
{\em formal} $O(\epsilon)$ perturbative result, (4.\ref{3.84}),
relies upon (\ref{3.92'}), not (\ref{3.92}). That is, if the
perturbation $f$ entails singularities there are deep potential
problems unless $\epsilon$ is viewed as an actual infinitesimal.

Now, what we might have liked to say is that the region of
concern is just $[0,\hat{\xi}]$, the domain of physical
fluid, so that even if $f$ has singularities outside this region,
we need pay attention only to a domain free of them in which
(\ref{3.92}) and (\ref{3.92'}) are equivalent. That {\em both}
interfaces must be satisfied, with its specific form of the method
of images, (4.\ref{3.34"}) prevents us from determining satisfaction
just within $[0,\hat{\xi}]$. By the equations of motion for
the driven interface, singularities within
$[-\hat{\xi},0]$ must be compatible (e.g. have residues
correctly determined to $O(\epsilon)$) with the analytic form of
$f_p$ within $[0,\hat{\xi}]$. But also, the simultaneous
satisfaction of the $g$ equation, yielding (4.\ref{3.34"}) requires
including the behavior of $f$ up to a least $2\hat{\xi}$, if
not even $3\hat\xi$ (since $-\zeta$, near a singularity,
can approach $\hat{\xi}$). Thus,
$f_p\simeq\hat{f}+\epsilon f$ must consistently agree with
singularity structure over a region basically
$[-\hat{\xi},3\hat{\xi}]$. Indeed there is every reason
to expect $\hat{f}+\epsilon f$ to arbitrarily disagree outside
this domain, so far beyond physical fluid.

Let us first see that the up-summed solution, stabilities with their
far-away singularities, is in conformity with (4.\ref{3.34"}) and
(\ref{3.92}) throughout this domain, even at singularities. Within
this domain, only the leading term (4.\ref{3.78}) possesses a
singularity, at $\zeta_s+2\hat\xi$ when $R(\nu)$ has a
singularity at $\zeta_s$. Since in this case
$\zeta_s\in [-\hat\xi,0]$ in the worst case (strongest
perturbation), $f_p$ must be analytic at $-\zeta_s$, which is
then in physical fluid. So, in (4.\ref{3.34"}) we consider
$$\zeta=-\zeta_s+x,$$
\bee
f_p(\zeta_s-x)-f_p(2\xi_g+\zeta_s-x)\sim\Gamma(f_p(-\zeta_s+x)) \nonumber \\
\sim\hat\Gamma(f_p(-\zeta_s+x))+\epsilon R(e^{\hat{f}(-\zeta_s+x)}).     \label{3.93}
\eee 

Now, $f_p$'s perturbing part is singular at $2\xi_g+\zeta_s$
by (4.\ref{3.78}), $f_p$ is regular at $\pm\zeta_s$, as is
$\hat{\Gamma}(f_p(-\zeta_s)))$, but $R$ is
also singular at $-\zeta_s$. Indeed, (4.\ref{3.78}) of perturbation
theory exactly agrees with this, and so, we conclude that
up-summing, even with singularities, is compatible with $f_p$ existing
as an approximate solution.

Next we turn to down-summing. With $R(\nu)=\nu^n$,
$n>0$, it is elementary to exactly sum (4.\ref{3.74}), while by
(4.\ref{3.72}), such a perturbation is flux-bearing with
$\Psi_n=-\lambda^{-n}=-e^{2n\varphi}$, rapidly exponentially growing, but
visible to the pump driving the flow, and so in principle,
controllable. These perturbations are nonsingular, and so for small
enough $\epsilon$ (\ref{3.92}) and (\ref{3.92'}) are equivalent. At the
end-points of the domain we must consider, the perturbing $f$'s
become enormous, so that $\hat{f}+\epsilon f$ is probably reliable only
for $\epsilon\sim e^{-L}$. For larger $\epsilon$, when the
perturbation has grown significantly with $\varphi$, we have no
idea what its nonlinear fate is.

Finally, we consider summing the $R_n$'s into
singularities, to produce, for example, pole-dynamical $R$'s
such as $\ln(1+a\nu)$. Let us now see that if these singularities
lie within $[-\hat{\xi},0]$, then the perturbative result is not
an approximation to any solution of the problem. The test is for
$$\zeta=\zeta_s+x$$
where, for $\zeta_s\in [-\hat\xi,0]$, $-\zeta_s$ is within
physical fluid. Here, (4.\ref{3.34"}) is
$$f_p(-\zeta_s-x)-f_p(2\xi_g-\zeta_s-x)=\Gamma(f_p(\zeta_s+x)),$$
or, with (4.\ref{3.91}), and picking off singular terms, 
\be
-\epsilon f_d(2\hat{\xi}-\zeta_s-x)\sim \hat{\Gamma}(\hat{f}(\zeta_s)+\epsilon
f_d(\zeta_s+x))+\epsilon R(\hat{f}(\zeta_s)). \label{3.94}
\ee
Now, $R$ is regular at $\hat{f}(\zeta_s)$ (it's singular at
$\hat{f}(-\zeta_s)$), but $\hat{\Gamma}$ is to be evaluated at an
infinite value of its argument, at which, in fact, it is regular.
However, by (4.\ref{3.74}), $2\hat{\xi}-\zeta_s$ {\em is} a singularity of
$f_d$, and indeed, within the domain of concern, only $\zeta_s$ and
$2\hat{\xi}-\zeta_s$ are singularities. But then, (\ref{3.94}) is
inconsistent. (Had we used the incorrect form, (\ref{3.92'}), the
two sides would have agreed with identical residues.)
{\em That is, down-summed solutions are compatible with
being approximations provided that} $R(\nu)$ {\em is free of
singularities in} $[-\hat{\xi},0]$. This implies isolated
pattern selection. To produce a finger of different width, the
leading pole-dynamic term $R(\nu)=\ln(1+a\nu)$ is required
for values of $a$ not exponentially small with fluid length $L$. In
this case during the finger regime $\lambda\ll 1$,
$k\lambda\gg 1$, or $1\ll\varphi\ll L$, the singularity
$\zeta_s$ must move $O(e^{-2\varphi})$ near to $\zeta=0$ for the
logarithmic stretching to take place that modifies finger width.
To say that $\hat{f}+\epsilon R$ for fixed finite $\epsilon$ is an
approximate solution is that
$$f_p\simeq\hat{f}+\epsilon\ln(1+a/\lambda)+\epsilon\ln(1-e^{\zeta_s-\zeta}),$$
so that near $\zeta_s$,
$$f_p'\sim\frac{\epsilon}{e^{\zeta-\zeta_s}-1},\;
f_{p,\varphi}\sim\frac{-\epsilon\zeta_s'}{e^{\zeta-\zeta_s}-1}$$
so that
$$\frac{f_{p,\varphi}}{f_p'}\overrightarrow{\zeta\rightarrow\zeta_s}-\zeta_s'.$$

Dividing the equation of motion for the driven interface by
$f_p'(\zeta)$, we have, just as for pole dynamics
$$f_p(-\zeta_s(\varphi),\varphi)=x_s+O(\epsilon^2), \; x_s'=0.$$
That is, if $\ln(1+a\nu)$ is to produce a changed finger, $f_p$
near Re$\zeta=0$ must include the singularity and $f_p(-\zeta_s)$ must produce a
stagnation point.  $f_p$ near its other nearest singularity, presumably at
$2\hat{\xi}-\zeta_s$ must be compatible with the method of images (4.\ref{3.34"}). That is,
(4.\ref{3.34"}) must be satisfied for $|{\rm Re}\zeta|<\delta$ with 
$\delta\rightarrow 0^+$.  
This is exactly what we have demonstrated is false in (\ref{3.94}).

Indeed, with singularities never within $[-\hat{\xi},0]$, the perturbation is 
exponentially small
with fluid length during the finger regime, and so for a long enough body of fluid,
without impact on the solution $\hat{f}$.  That is, we have ruled out the possibility
raised at the beginning of Section 4. that requesting an exactly translating solution
might be too brittle a question to determine selection for a finite body of fluid.  
Furthermore, with $\hat{f}$ warding off singularities in $[-\hat{\xi},0]$, or roughly 
by the length
of physical fluid, the greatest host of perturbations that could induce finite-time
singularities have also been warded off.  All that is significantly left are the
exponentially growing $\nu^n$ perturbations.  Left to themselves, these rapidly lead
$\hat{f}$ into a nonlinear regime about which we know little, but surmise 
tip-splitting.  On the other hand, as they proportionately modify flux, they are
controllable.  It is conceivable that a simulation of this flow constrained to prevent
rapid flux variations might, near enough to $\hat{f}$, be stable.

Apart from these exponentially growing fluxes, there are no slow modes to modify
$\hat{\xi}\sim L_0-\varphi$ to $\hat{\xi}\sim L_0-(1/\lambda-1)\varphi$.  That is, 
this flow protects the far downstream behavior of the 1/2 finger
without surface tension on the efflux.  The ability of $\hat{f}$ to ward off nearby
singularities lies in its ability to satisfy (4.\ref{3.34"}) {\em without} any image 
singularities present, whereas all putative nearby perturbations with singularities 
must possess them, and are then inconsistent.  Clearly, with surface tension present,
a method of images reminiscent of (4.\ref{3.34"}) must exist, and so one can surmise
that the appropriate S-T solution is again special.
Even though surface tension is a perturbation
that spoils the critical symmetry of this paper, there is now a reinforced reason to 
surmise that (3.\ref{2.36}),
$$1-2\lambda=(2\pi)^2 B_e,$$
at least for $\lambda$ near 1/2 is correct, which I offer as a conjecture.


\section{Precis and Discussion}

We have followed a clue laid down by reflection symmetry that the analytically 
continued equations of motion are symmetric under parity, $\zeta\rightarrow -\zeta$.
This suggests 
that far upstream (${\rm Re} \zeta<0$) singularities, which determine the shape of the 
driven 
interface at Re$\zeta=0$, are related to and co-determined with far downstream behavior,
which is the nature of the efflux in a physical experiment.  To see if any such 
thing holds, it was then mandatory to consider the flow of a finite body of fluid. 
The simplest feasible such problem, fully posed within the purview of the 
analytically continued conformal machinery, consists in replacing the Riemann 
mapping of a disc to an annulus.  The case of the simpler disc embraces all prior 
studies, and corresponds to fluid going off infinitely far to the right with a 
simple pole at $+\infty$.  In the annular version, the available channel is still infinite,
but fluid terminates at a downstream interface, the behavior of which constitutes 
the efflux for this configuration.

Requesting a purely translating driven interface receives a unique solution in the 
finite problem:  There is sharp, precise selection for this question.  Examined in 
pole-dynamics, even including singularities to the right of the efflux interface, 
precisely and only this selected solution exists.  Examining a class of solutions 
that exist for the problem of the disc, but not here; namely the usual 
Saffman-Taylor solutions, we notice that selection is expressed  on the efflux 
interface, even when asymptotically far downstream, with only the 1/2 solution 
maintaining it at a fixed pressure.  Solutions with $\lambda<1/2$ also correspond to 
solutions to the finite problem, but with surface tension now smoothing the efflux 
interface (although, of course, with zero surface tension still on the driven 
interface).  The explicit relationship, $1-2\lambda=\sigma/v$ provides the physical 
interpretation of all the S-T solutions.  Moreover, that the infinite fluid version 
of the disc allows all these possibilities with equanimity, determines that its 
pole at $+\infty$ is truly within physical fluid, so that the problem of the disc 
is the 
theory of an unfinished, unterminated experiment, and is not the limit of a very 
long body of fluid.  That is, the physical limit $L\rightarrow \infty$ is singular.

We end up showing that solutions to the finite problem near the selected one all 
preserve the efflux condition $1-2\lambda=0=\sigma/v$. Although we have not worked out the 
theory for surface tension on the efflux, considering that the $\lambda\neq 1/2$ 
solutions do 
meet this condition, and that the condition is conserved under perturbations in 
the $\lambda=1/2$ case, it is the most natural surmise that the width-efflux condition is 
generally correct, at least for $\lambda$ not too much below 1/2.

Most importantly, we reach the conclusion that while many consequences of the 
distant efflux are exponentially small, this is false for pattern selection.  
This is so different from the generally presumed behavior that for this reason 
alone, the annular configuration we discuss should be put to the experimental test.
It is important in this regard to note (private communications from several of the 
experimenters) that this selection with beautifully formed fingers is not so easy 
to experimentally produce, generally requiring fiddling with the efflux to become 
reliably reproducible.  For example, Tabeling {\em et al}\cite{87TZL} succeeded only 
after fitting 
an ``impedance-matching" plug to the efflux. It is regrettable that the 
experimenters haven't carefully observed and calibrated such dependencies.  
This is not for any failure of the experimenters, but rather, illuminates an 
important aspect of the relationship between theory and experiment.  Namely, 
unless one has prior insights - basically theoretical knowledge - then one doesn't 
quite know what is worthy to observe.  Necessarily, in order to better observe 
what one has chosen to, the experimenter exercises his art to control what 
obfuscates.  But some of these thrown-out details might be at the actual heart of 
the system being observed.  How is one to know in advance? And certainly, until 
certified by experiment, there is no way to be sure that the theory considered is 
significantly related to the physical phenomenon.

Having determined that at least for the elementary S-T solutions that only 
$\lambda=1/2$
has any connection to a physically finite flow with efflux at atmospheric pressure 
and no surface tension on it, we go on to question if asking for pure translation 
in a finite problem is perhaps too brittle a request.  In the flow of a finite 
piece of fluid, there is a finite, definite period of time over which the solution 
is that of a finger with a large body of fluid still downstream from it.  The 
driven interface at velocity $1/\lambda$ penetrates into the fluid whose terminating 
interface, virtually flat, moves with velocity 1.  For $\lambda=1/2$, by the time the 
fluid has translated its resting length $L$ under the mean velocity of its flux (i.e. 
1), there remains but a thin layer of fluid still downstream from the driven 
interface, as the efflux interface begins to curve into a similar finger.  Calling 
this finite period of time when the pure finger is well-formed (i.e. with long 
horizontal sides) but still with a large body of fluid behind it the ``finger 
regime" (as opposed to afterwards, the ``breakthrough regime"), we can ask if there 
are solutions that within measurable error appear as perfectly translating, but 
actually possess deviations exponentially small with $L$, only becoming physically 
important in the breakthrough regime.  Should this be the case, then the strong 
selection is a misleading artifact of too mathematically posed a question.

Since there are no exact solutions we know of (e.g. all versions of pole-dynamics) 
save for 
the pure 1/2 S-T finger, we regrettably had no choice but to perform a perturbative
analysis.  As we erect this machinery, we are immediately led to notice that 
exceptionally useful new variables, basically 
$(f(\zeta,\varphi),f(-\zeta,\varphi))$, constructed from 
the old $(\zeta,\varphi)$, are guaranteed of invertiblity, since the equation of motion for 
the driven interface is precisely the statement that the Jacobian is the constant 
2.  Translating in $\zeta$ an identical fact for the equation of motion for the efflux 
interface, we notice that given the first equation of motion, the second can be 
fully integrated into the relationship 
\[ \hspace{10mm}
f(-\zeta,\varphi)-f(2\xi_g-\zeta,\varphi)=\Gamma(f(\zeta,\varphi)). \hfill 
(\Gamma)
\]
What is crucial here is that $\Gamma$ depends {\em only} upon the one new variable 
$f(\zeta,\varphi)$.  This 
formula serves as the precise way in which the method of images is to provide a 
solution compatible with both interfaces.

In these new variables (more conveniently, their exponentials, $(\xi,\nu)$) perturbation 
theory is especially simple for just one interface.  Together with $(\Gamma)$ expressed 
perturbatively, we are able to fully integrate the perturbed p.d.e.'s for both 
interfaces, and reduce the problem to a purely algebraic one.  We explicitly did 
so here for the first two orders of perturbation, and to first order wrote down 
all possible solutions.  The most striking such solutions are just the usual 
infinite fluid pole dynamical ones, but decorated with exponentially small image 
terms, vanishing with $L\rightarrow \infty$.  Since we already rigorously know 
that there are no 
such solutions, we wonder how they perturbatively appear so naturally and 
fluently.  Before discussing this conundrum, we notice something central to the 
heart of the finite problem.

A finite body of viscous fluid has the natural constituentive relationship of 
impedance, the ratio of the net pressure across the fluid to the flux of its 
transport.  Although impedance is infinite in the usual infinite fluid problem, it 
is less clear that its time derivative should also exactly vanish, as it does in 
the theoretical literature, but ostensibly in disagreement with experiment.  
This is a signature that the infinite boundary geometry is 
unphysical, even construed as the limit $L\rightarrow \infty$.  Should a deformation of the 
fluid's flow produce a change in impedance, that is to say a change in the 
conserved flux through the entire fluid when pressure across it is maintained 
fixed, then such a deformation requests power from the pump that maintains this 
pressure difference.  That is, such a deformation is {\em driven} by the pump, is 
controllable by the pump, and is certainly not spontaneously impressible.  
Such a circumstance is hardly surprising since a deformation is not just on the 
driven interface, but throughout the incompressible fluid that must everywhere 
adjust to accommodate to a distortion of its boundary.  This is very different in 
the infinite fluid geometry, where the promiscuous pole at infinity simply picks 
up whatever need be, since for the Riemann disc, the disposition of the one 
interface uniquely determines everything.

Performing the perturbative analysis of the finite fluid, we are forced to 
immediately decide if the perturbation is to modify flux, and so be externally 
driven.  Should this not be allowed, we discover that the unperturbed finger has 
only purely stable modes, although not enough to allow for the arbitrary 
independent perturbations of both interfaces.  The residual modes, all 
instabilities, are all driven by the external pump, and so, in no sense free 
fluctuations.  These same modes, as $L\rightarrow \infty$, are precisely all the 
modes of the infinite fluid problem, although in that configuration they are all
free.  
This now cements the observation that $L\rightarrow \infty$ is singular,
and the infinite fluid geometry a physically wrong limit.  There is no reason to 
doubt that these exponentially growing modes are real - they are, however, 
controllable. Indeed, a record of the flux through, or the power delivered by the 
pump over the course of time uniquely determines the disposition of this unstable 
flow throughout the entire body of fluid, and throughout this interval of time.

Granting these modes and an experimental control that allows them to grow at most 
algebraically in time, we now ask about the slow modes produced by summing all 
these fast modes.  We especially care about summing them into logarithms, which 
then modifies the linear change of impedance in time, which then modifies the 
efflux transport from $L-\varphi$ to $L-(1/\lambda -1)\varphi$, hence modifies 
$\lambda$ from 1/2, and so, with 
no surface tension on the efflux produces a flow that violates 
$1-2\lambda=\sigma/v$.
Here, 
resorting to $(\Gamma)$, we discover that these logarithmic modes are in fact purely 
formal, where with $\epsilon$ the strength of perturbation, they fail to be 
solutions for 
any finite value of $\epsilon$, no matter how small, thus resolving the conundrum that they 
don't exist in any correct non-perturbative analysis.  With this, we establish 
that the solution of $\lambda=1/2$ is definite and isolated in the neighborhood of the 
pure unperturbed finger, and that we had not posed too brittle a question in asking
for exact translation.  Evidently, one would far prefer a purely non-perturbative 
treatment for a problem with a rather subtle resolution.  Nevertheless, this work 
serves as a tonic against the use of linear response methods in such delicate 
systems.


\section{Appendix: Selection Within Pole Dynamics}
\setcounter{equation}{0}

The $\varphi$-translation invariant solutions of (3.\ref{2.50}), which we shall here
refer to as $\hat{f}_n$ with $n$ as in that equation, belong
to a certain well-studied class of solutions (Class (27) of Reference \cite{1} )
We used no properties of that Class, but rather deduced (3.\ref{2.50})
from dynamical symmetry arguments.  This Class
has a most important property: its members are exact
solutions of $(2.14')$. These particular dynamical solutions  
are the so-called ``pole dynamics''. We shall, to render this 
paper more self-contained, re-derive these results here, as they are so
trivial under reflection-symmetry, but then easily extend them to more 
complex variants.
Most importantly, we shall see that even within this significantly enlarged
class of pole dynamics, the {\em only} solutions for both interfaces 
within the class are precisely the $\hat{f}$'s of (3.\ref{2.50}).  That is, there exist
no perturbations to $\hat{f}$ whatsoever in this general class, which is 
to say that had the dynamics been exhausted by pole dynamics, then $\hat{f}$ is
rigidly isolated as {\em the} solution, and hence sharp pattern selection.

We start with the delineation of pole dynamics (Class(27)): \\
\begin{eqnarray}
f=\beta(\varphi)+\zeta+\Sigma\alpha_k \ln(1-e^{\zeta_k-\zeta});  \label{a.53} \\ 
{\rm Re} \zeta_k(\varphi)<0; \; \; \alpha_k=  {\rm{const}} \nonumber
\end{eqnarray}
where $f$ must satisfy $(2.14')$ and be reflection-symmetric ($\beta$ real;
$(\alpha_k,\zeta_k), (\bar{\alpha}_k, \bar{\zeta_k})$ both present).
Paying attention to $(2.14')$, write
\begin{equation}
f'=1 + \Sigma\frac{\alpha_k}{e^{\zeta-\zeta_k}-1}~;~ f_{\varphi}=
\beta'-\Sigma\frac{\alpha_k\zeta'_k}{e^{\zeta-\zeta_k}-1} \label{a.54}
\end{equation}

Consider $\zeta \rightarrow \zeta_k$. Then $1/f'(\zeta) \rightarrow 0, \; \;
f_{\varphi}/f'(\zeta)\rightarrow -\zeta'_k(\varphi)$.  Dividing $(2.14')$ by
$f'(\zeta)f'(-\zeta)$ and taking $\zeta \rightarrow \zeta_k$,
\begin{eqnarray}
0 & = & -\zeta'_k+f_\varphi (-\zeta_k)/f'(-\zeta_k); \nonumber \\
0 &=& f_\varphi(-\zeta_k,\varphi)
-\zeta'_k (\varphi)f'(-\zeta_k,\varphi)=\frac{d}{d\varphi}
f(-\zeta_k,\varphi) \nonumber \\ 
& \Rightarrow & f(-\zeta_k,\varphi)=\bar{z}_k= \; \; {\rm{const}}. \label{a.55}
\end{eqnarray}

These points, $z_k=f(-\bar{\zeta}_k, \varphi)$ are ``stagnation'' points,
in the sense that as $\varphi \rightarrow \infty$ all Re $\zeta_k
\rightarrow 0^-$, so that $z_k$ is very near to a point on the interface
$\zeta=i s$ , and the interface may never pass through this point. More 
correctly, the interface must asymptotically come to rest at
\begin{equation}
z^*_k = z_k-\alpha_k \ln 2 \label{a.56}
\end{equation}
since as Re $\zeta_k \rightarrow 0 ~ -\bar{\zeta}_k \rightarrow \zeta_k$,
and $f$ is singular at $\zeta_k$.  Setting $\zeta=i\eta_k+it\xi_{k}$ with
$\zeta_k=\xi_k+i\eta_k$, writing down $f(\zeta), f(-\bar{\zeta}_k)$, 
and subtracting yields
\be
f(i\eta_k+it\xi_k) \sim z_k-\alpha_k \ln 2+\alpha_k \ln (1-it). \label{a.56'}
\ee
For a flow within an initially almost flat interface that grows wrinkled,
all the $z_k$ have real parts to the right of the interface, and so, by (\ref{a.56})
must have Re $\alpha_k > 0$, since the interface can never have passed 
through $z_k$ itself.

(\ref{a.55}) determines all the $\zeta_k(\varphi)$ in terms of $\beta$. $f$ has 
derivatives analytic at Re $\zeta \rightarrow \pm \infty$: By (\ref{a.54}),
\begin{eqnarray}
f'(+\infty)&=&1, \; \; ~  f'(-\infty) =1-\Sigma\alpha, \nonumber \\
 f_\varphi(+\infty)
&=& \beta',\; \; f_\varphi(-\infty)=\beta'+\Sigma\alpha_k\zeta_k', \label{a.57}
\end{eqnarray}
and so by $(2.14')$,
\begin{equation}
(1-\Sigma\alpha)\beta+(\beta+\Sigma\alpha_k\zeta_k)=
2\varphi+ ~ {\rm{const}}, \label{a.58}
\end{equation}
thus fully determining $f$.

As $\varphi \rightarrow \infty \; \zeta'_k \rightarrow 0$, as Re $\zeta_k 
\rightarrow 0^-$, and by $f$'s imaging, Im $\zeta_k \rightarrow 0, \pi$
only (see Paper I).  By (\ref{a.54}) outside arbitrarily small disks about the 
$\zeta_k, f_\varphi \sim \beta'$, and by $(2.14')$,
$$\frac{2}{\beta'}\sim f'(\zeta)+f'(-\zeta)$$
with $f'_\varphi\sim 0 $, so $f'(\zeta,\varphi) \sim f'(\zeta)$. Hence
\begin{equation}
\beta ' \equiv \frac{1}{\lambda}= \;  {\rm{const}}, \label{a.59}
\end{equation}
and $f'(\zeta)+f'(-\zeta) \sim 2\lambda, f(\zeta)-f(-\zeta)\sim 2\lambda\zeta$
and so on $\zeta=i s$,
\begin{equation}
 y(s)={\rm Im} f(i s)\sim \lambda s
\; \; \; s \in (\epsilon, \pi-\epsilon) \label{a.60}
\end{equation}
which is to say it is a finger of width $\lambda$ of the channel. By (\ref{a.58})
and (\ref{a.59})
\begin{equation}
\lambda - \frac{1}{2}=\frac{1}{2}(1-\Sigma \alpha)=\frac{1}{2}f'(-\infty).
\label{a.61}
\end{equation}

Now, the $\hat{f}_{n}$ of (3.\ref{2.50}) are explicitly of form (\ref{a.53}) with 
$f'(-\infty)=0$, 
and so $\lambda=1/2$.  We ultimately care about $\hat{f}_1$, a single 1/2 finger in
a $2\pi$ channel.  It is now natural to ask if there are other Class (27)
solutions for both interfaces, since then we could analytically solve the 
problem of such perturbations to $\hat{f}$.  It is easy to see that there
are none.

First, substitute (2.\ref{15}) into $(2.14'')$ to obtain
\begin{eqnarray}
2-f'(-\zeta+\xi_g)f_\varphi(\zeta+\xi_g)-f'(\zeta+\xi_g)f_\varphi(-\zeta+
\xi_g) \nonumber \\
= 2\xi'_g f'(\zeta+\xi_g)f'(-\zeta+\xi_g). \label{a.62}
\end{eqnarray}
with $\xi_g$ finite, taking the limit Re $\zeta \rightarrow +\infty$,
where Class(27) is analytic, the left hand side vanishes, as it is the 
limit of simultaneous $(2.14')$. Hence the only $f$'s that are allowed satisfy
\be
\xi'_g(\varphi) f'(+ \infty) f'(-\infty) = 0 \label{a.62'}
\ee
But $f'(+\infty)=1$, and for any non-flat solution $\xi'_g\neq 0$, and so 
$f'(-\infty)=0, \lambda=1/2$, and so,
\be
\Sigma \alpha = 1 \label{a.62''}
\ee
is a consistency condition for $f$ to obey both equations of motion.  Thus,
$\lambda=1/2$ is selected.

There is an important simple fact related to why $\xi'_g\neq 0$.  There is a unique
conformal ({\em i.e.} invertible) map $h$ from the fluid at time $\varphi$ to a rectangle
with sides $\xi_g$ and $2\pi$.  The aspect ratio, $\xi_g / 2\pi$, of this rectangle is 
termed the ``module" of the region mapped, and is uniquely determined by that region.
Moreover, it enjoys an exact estimate:
$$2\pi \xi_g(\varphi) \le A(\varphi) \equiv A_0$$
with $A$ the area of the fluid, the constant in time $A_0$.  Unless $h' \equiv 1$, the 
inequality is strict.  Thus, with $A_0 \equiv 2\pi L_0$, we have
$$\xi_g(\varphi) \le L_0.$$
That is, for other than the flat flow $\xi_g(\varphi)$ is stricly smaller than the resting
length of the fluid.  In particular, for a solution which was flat in the far past,
$\xi'_g\neq 0$, since it had the value $L_0$ in the far past, but certainly below it at
finite times.

With $f$ of (\ref{a.53}), $g$ of (2.\ref{15}) is
\be
g=(\beta+\xi_g)+\zeta+\Sigma \alpha_k \ln (1-e^{\zeta_k-\xi_g-\zeta}). \label{a.53'}
\ee
Between $f$ and $g$, the $\zeta \rightarrow \infty$ equation, (\ref{a.58}) is identical 
with $\Sigma \alpha=1$.  So, there is one such equation for $\beta$.  However
together with (\ref{a.55}), we now also have
\be
g(\xi_g-\bar{\zeta}_k,\varphi)=z^+_k. \label{a.55'}
\ee
With a real $\zeta_k$, by $f$'s reflection symmetry such a (\ref{a.55}), 
$(\ref{a.55'})$
is a real equation.  For each complex $\zeta_k$, each is complex, and hence a 
pair of real equations.  So, with $m$ complex $\zeta_k$'s and $n$ real ones, 
we have
\begin{equation}
{\rm{\# \;  real \;  eq's}}  \; = 4m + 2n + 1 \label{a.63}
\end{equation}
and
\be
{\rm{\# \;  real  \; vbl's}}  \; = 2m + n + 2 \; \; 
(\zeta_k {\rm{'s}} \; \; +\xi_g+\beta) \label{a.63'}
\ee
and so the system is overdetermined by
\be
{\rm{excess}}  \; = 2m+n-1. \label{a.63''}
\ee
So, save for the case of just one $\zeta_k \equiv \hat{\zeta}$ which is 
real, and even with just one complex $\hat{\zeta}$, the system is 
overdetermined. The one real case is just $\hat{f}_1$ of (3.\ref{2.42}).

Let us now show that the only solutions that become flat as $\varphi
\rightarrow -\infty$ are just the $\hat{f}_n$, with the overdetermination
in fact inconsistent in all other cases.

Writing down (\ref{a.55}) for a $\zeta_k$ and the difference of $(\ref{a.55'})$ 
and (\ref{a.55}),
we have 
\begin{equation}
\bar{z}_k=\beta-\zeta_k+\Sigma \alpha_\ell \ln (1-e^{\zeta_l+\zeta_k}) \label{a.64}
\end{equation}
\be
\bar{z}^+_k -\bar{z}_k \equiv \Delta_k=2\xi_g + \Sigma \alpha_\ell \ln
\left(\frac{1-e^{-2\xi_g+\zeta_k+\zeta_\ell}}
{1-e^{\zeta_k+\zeta_\ell}}\right). \label{a.64'}
\ee
For all Re $\zeta_k \rightarrow -\infty, (\varphi \rightarrow -\infty),
\xi_g \rightarrow \xi_0$ (finite), we immediately have by $(\ref{a.64'})$
\begin{equation}
\Delta_k \equiv 2\xi_0. \label{a.65}
\end{equation}
Expanding $(\ref{a.64'})$, since all Re $\zeta_k < 0$ and $\xi_g > 0$ and
so the series are absolutely convergent, produces 
$$\Sigma \alpha_\ell \sum^\infty_{r=1}\frac{1}{r}(1-e^{-2r\xi_g}) 
e^{r(\zeta_k+\zeta_\ell)}$$ which can be resummed as
$$\sum^\infty_{r=1}\left[\frac{1}{r}(1-e^{2r\xi_g})
(\Sigma \alpha_\ell e^{r\zeta_\ell})\right]e^{r\zeta_k}.$$
So, 
\begin{eqnarray}
2&(&\xi_0-\xi_g)=\sum^{\infty}_{r=1} u_r(\varphi)e^{r\zeta_k};\label{a.66} 
\\ \nonumber
u_r&(&\varphi)\equiv\frac{1}{r}(1-e^{-2r\xi_g})\Sigma\alpha_\ell
e^{r\zeta_\ell}.
\end{eqnarray}
The first of (\ref{a.66}) implies that
\begin{eqnarray}
u_r(\varphi)&\equiv& 0 ~ \; r<n; \; u_n \neq 0 \nonumber \\
{\rm{and}} \; 
e^{n\zeta_k} &\equiv& e^{n\zeta_o} \; \; \; {\rm{for \; some}} \; \; \;
n\geq 1. \label {a.67}
\end{eqnarray}
Thus
\begin{equation}
\zeta_k=\zeta_0+2\pi i \frac{k}{n} \; \; \; k=0,...., n-1. \label {a.68}
\end{equation}
For $u_{r<n}\equiv 0$, by the second of (\ref{a.66}), 
$\sum^{n-1}_{\ell=0} \alpha_\ell e^{2\pi i \frac{r \ell}{n}}=0 \; ,
 r=1,....,n-1$,
and so by finite Fourier, with $(\ref{a.62''})$, $$\alpha_k \equiv 1/n.$$

Finally, taking Re $\zeta$ sufficiently large and analytically continuing
the result,
\begin{eqnarray}
f & = & \beta+\zeta+\frac{1}{n}\sum^{n-1}_{k=0} \ln (1-e^{\zeta_0-\zeta} 
e^{2\pi i k/n})  \nonumber \\
& = & \beta+\zeta-\sum^\infty_{r=1}\frac{1}{r}e^{r(\zeta_0-\zeta)}
\left(\frac{1}{n}\sum^{n-1}_{k=0} e^{2\pi i k r/n} = 
\sum_\ell \delta_{r,n\ell}\right) \nonumber \\
& = & \beta+\zeta- \frac{1}{n}\sum^\infty_{\ell=1}\frac{1}{\ell}
e^{n \ell(\zeta_0-\zeta)} \nonumber \\
&=& \beta+\zeta+\frac{1}{n} \ln (1-e^{n(\zeta_0-\zeta)}). \label{a.69}
\end{eqnarray}
By reflection-symmetry $e^{n\zeta_0}$ is real, or $e^{n\zeta_0}=
\pm e^{n\hat{\zeta}}$ for $\hat{\zeta}$ real, and so
$$f=\beta+\hat{\zeta}+\frac{1}{n}\ln(e^{n(\zeta-\hat{\zeta})}\pm 1)$$
which then to satisfy $(2.14')$ is just $\hat{f}_n$ of (3.\ref{2.50}).  It must be
remembered that $(2.14')$ is nonlinear: the only $f$'s of Class (27) is
precisely \underline{one} of these $\hat{f}_n$'s \underline{not} any
linear combinations whatsoever.  That is, there can be \underline{no}
Class (27) perturbations to $\hat{f}$, so that within this Class, 
$\hat{f}$ is perfectly stable of unique 1/2 width.

Now, Class (27) arose in discussions of the infinite channel: in the 
natural variable $e^{-\zeta}\equiv\omega, u=e^{-f}$ is analytic at
$\omega=0$, where it has a simple zero. $e^{-f}$ is the unique (under
reflection symmetry) Riemann map taking $u=0$ to $\omega=0$ of the
physical region to the unit disk.  With two free boundaries, the map
is to the annulus $e^{-\xi_g}<|\omega|<1$, and so is generally not
analytic in $|\omega|<e^{-\zeta_g}$. The $f$'s of (\ref{a.55}) are a Taylor 
expansion about $\omega=0$. For two free boundaries, it is generally to be
a Taylor-Laurent expansion.  Let us write this as
\bee
f=\beta+\zeta &+& \Sigma \alpha_k \ln (1-e^{\zeta_k-\zeta})-\Sigma \alpha_k^+ 
\ln (1-e^{-\zeta^+_k +\zeta}) \nonumber \\
& {\rm with} & ~~ {\rm Re} \zeta_k<0,~ {\rm Re} \zeta^+_k>0. \label{a.53^+}
\eee

To fully consider these solutions, we extend the pole-dynamics
of (\ref{a.55}) to include the special circumstance that the $\zeta_k$ and $\zeta^+_k$
might be matched with $\zeta^+_k\equiv -\zeta_k$, which is the ``closest''
that $(\ref{a.53^+})$ can come to a solution (There are none.) 

Writing the derivatives,
\bee
f' &=& 1+\Sigma\frac{\alpha_k}{e^{\zeta-\zeta_k}-1}+\Sigma\frac{\alpha^+_k}
{e^{\zeta^+_k-\zeta}-1}; \nonumber \\
f_\varphi &=& \beta'-\Sigma\frac{\alpha_k{\zeta_k}'}
{e^{\zeta-\zeta_k}-1}-\Sigma\frac{\alpha^+_k{\zeta^+_k}'}
{e^{\zeta^+_k-\zeta}-1}. \label{a.54'}
\eee

First
\bee
f'(+\infty)= 1-\Sigma\alpha^+ &,& f'(-\infty)=1-\Sigma\alpha, \nonumber \\
f_\varphi(+\infty)
=\beta'+\Sigma\alpha^+_k {\zeta^+_k}' &,& f_\varphi(-\infty)=\beta'+\Sigma\alpha_k
\zeta_k', \label{a.57'}
\eee
so that $(2.14')$ yields
\bee
(1-\Sigma\alpha)(\beta+\Sigma\alpha^+_k\zeta^+_k) &+& (1-\Sigma\alpha^+)
(\beta+\Sigma\alpha_k\zeta_k) \nonumber \\ 
&=& 2\varphi+ \; {\rm{const.}} \label{a.58'}
\eee
It is next immediately clear that (\ref{a.55}) is unmodified for both $\zeta_k$ 
{\em{and}} $\zeta^+_k$ {\em{if}} they are unpaired:
\be
f(-\zeta_k,\varphi)=\bar{z}_k, f(\zeta^+_k,\varphi)=\bar{z}^+_k,
\zeta_k+\zeta^+_k \neq 0. \label{b.55'}
\ee
Should $\zeta_k+\zeta^+_k=0$ (arrange the indices so that mating $\zeta$'s 
have the same index), then the derivation of (\ref{a.55}) fails.  Instead we compute
with $(\ref{a.54'})$ $f'(\pm(\zeta_k+\epsilon))$ and 
$f_\varphi(\pm(\zeta_k+\epsilon))$,
and proceeding with a little care determine that
\be
\bar{z}_k=\alpha_k f_R(-\zeta_k,\varphi)+\alpha^+_k f_R(\zeta_k,\varphi)
\label{a.55''}
\ee
where $f_R$ means drop the term that diverges at the specified argument of
$f$ {\em{precisely}} of form $(\ref{a.53^+})$.  Most importantly, there is just 
{\em{one}} equation $(\ref{a.55''})$ for the matched pair $\pm \zeta_k$ which 
lowers the overdetermination count of $(\ref{a.63''})$.

The problem is that it doesn't lower the count enough, since if a pair
matches in $f$ it can't in $g$ and visa versa: The singularities of $g$,
by $(\ref{a.53^+})$ are at $\zeta_k-\xi_g$ and $\zeta^+_k-\xi_g$, so that if $\zeta_k$
and $\zeta^+_k$ are paired, and hence with mean 0, in $g$ they have mean
$-\xi_g$ and are no longer matched, and so each produces its own equation
$(\ref{a.55'})$. This, together with the notion of the ``method of images'', 
determines
the best possible way to place $f$'s singularities.  Consider a real 
\begin{equation}
\zeta_0\equiv \hat{\zeta} <0. \label {a.70}
\end{equation}
With $-\hat{\zeta} >0$ and approaching 0 as $\varphi \rightarrow + \infty$,
it is almost impossible for $-\hat{\zeta}$ not to be in physical fluid,
and so take it as unmatched. Next consider more $\zeta_k$:
\be
\zeta_k=\hat{\zeta} -2k \xi_g \; \; \; k=0,....,n. \label{a.70'}
\ee
But then
\be
\zeta^+_k\equiv -\zeta_k = -\hat{\zeta} + 2k \xi_g \; \; \; k=1,....,n
\label{a.70''}
\ee
are all matched, greater than $2\xi_g$, so certainly outside physical fluid.
These are $f$'s singularities. For $g$, we have
\be
\zeta_k^g \equiv \zeta_{k-1} -\xi_g = \hat{\zeta}+\xi_g -2k \xi_g \; \; 
k=1,....,n+1
\label{a.70'''}
\ee
and
\be
\zeta^{+g}_k=\zeta^+_k -\xi_g=-\hat{\zeta}-\xi_g+2k\xi_g=-\zeta^g_k \; \;
k=1,....,n. \label{a.70^*}
\ee
Again all the $\zeta^g_k$ are matched, save for $\zeta^g_{n+1}$.  Thus
both $f$ and $g$ have $n$ matched pairs and one unmatched singularity.
(This is the best that can be arranged within the $\xi_g$ annular domain.)
Counting, this is $2n+3$ real equations for $2n+1$ singularities,
$\xi_g$ and $\beta$.  This has a right sense to it, save for the difficulty
that the $2n+1$ singularities are just combinations of $\hat{\zeta}$ and
$\xi_g$, so that there are precisely always 3 variables, and so, overdetermined
save for $n=0$, which is simply $\hat{f}_1$.  It should be reasonably clear 
that $f$ of $(\ref{a.53^+})$ affords no exact solutions other than the $\hat{f}_n$ 
of $\varphi$-translation invariance.  That is, there is \underline{no} pole
dynamics at all. Rather, that class, natural to the dynamics, truly arises
as a guessed-at extension of the Saffman-Taylor stationary solution, but 
exists only for 1/2 fingers in consequence of displacement invariance. 

That is, the 1/2 finger has unlimited stability and selection within the
scope of any conceivable pole dynamics.  Insofar as the finite-time 
singularities in the literature are all of pole dynamics, $\hat{f}_1$ is
evidently totally free from them.

\acknowledgments

I wish to thank Michael Shub and Boris Shraiman for valuable discussions during the 
assembly of this work.
I thank Itamar Procaccia for his useful critique of this present work, as well as for the 
many
discussions that led me into this work, and the collaboration of our previous paper, 
Ref.\cite{1}.  Finally, I am indebted to Leo Kadanoff for his detailed assistance in
heping me to better present what is written here.  Needless to say, the execution and
responsibility lies just with me and my ability to profit from his generous aid.

\end{document}